\RequirePackage[2020-02-02]{latexrelease}
\documentclass[aps,prb,twocolumn,superscriptaddress,amsmath,amssymb]{revtex4-2}

\usepackage[version=4]{mhchem}
\usepackage{graphicx}

\begin{document}

\title{Full superconducting gap in the candidate topological superconductor \ce{In_{1-x}Pb_{x}Te} for x = 0.2}
\author{M. P. Smylie}
\affiliation{Department of Physics and Astronomy, Hofstra University, Hempstead, New York 11549, USA}
\affiliation{Materials Science Division, Argonne National Laboratory, 9700 S. Cass Ave., Lemont, Illinois 60439, USA}

\author{Kaya Kobayashi}
\affiliation{Research Institute for Interdisciplinary Science, Okayama University, Okayama 700-8530, Japan}
\affiliation{Department of Physics, Okayama University, Okayama 700-8530, Japan}

\author{J. Z. Dans}
\author{H. Hebbeker}
\affiliation{Department of Physics and Astronomy, Hofstra University, Hempstead, New York 11549, USA}

\author{R. Chapai}
\author{W.-K. Kwok}
\author{U. Welp}
\affiliation{Materials Science Division, Argonne National Laboratory, 9700 S. Cass Ave., Lemont, Illinois 60439, USA}

\begin{abstract}
    High-pressure synthesis techniques have allowed for the growth of samples on the indium-rich side of (Pb,In)Te, which have increased superconducting transition temperatures compared to lead-rich compounds. In this study we present measurements of the temperature dependence of the London penetration depth $\Delta \lambda(T)$ in the compound \ce{In_{0.8}Pb_{0.2}Te}, which shows a bulk $T_{c,onset}$ of $\sim4.75$ K. The results indicate fully gapped BCS-like behavior, ruling out odd-parity, topologically nontrivial $A_{2u}$ pairing; however, odd-parity $A_{1u}$ pairing is still possible. Critical field values measured below 1 K and other superconducting parameters are also presented.
\end{abstract}

\maketitle

\section{Introduction}
Following the theoretical prediction \cite{Fu-TI-prediction, Zhang-Bi2Se3-predict} and subsequent discovery of bulk topological insulators such as \ce{Bi_2Se_3} \cite{XiaBi2Se3-discovery, HsiehBi2Se3-discovery, Mazumder-Bi2Se3} in which time-reversal symmetry and parity lead to a finite $Z_2$ topological index, the first topological crystalline insulator, SnTe \cite{Hsieh-SnTe,Tanaka-SnTe} was discovered, where the gapless surface states are  protected by crystalline mirror symmetry and a finite mirror Chern number.
It was found \cite{Hor-CBS, Qiu-NBS, Liu-SBS, Zhong-PbSnInTe-superconductivity, Erickson-SnInTe, Mizuguchi-SnAgTe} that doping these topologically nontrivial insulating compounds can result in bulk superconductivity.
ARPES measurements on both superconducting \ce{M_xBi_2Se_3} (M = Cu, Sr, Nb) and on superconducting \ce{Sn_{1-x}In_xTe} reveal that the doped materials still possess spin-polarized topologically protected surface states near the Fermi energy \cite{Polley-SnInTe-ARPES, Sato-SnInTe-topological, Wray-CBS-ARPES,Neupane-SBS-ARPES}, raising the potential for discovery of topological superconductivity.
Indeed, a unique nematic topological superconducting state \cite{Yonezawa-Bi2Se3-review, FuBerg-CBS, FuCBS} was discovered in \ce{M_xBi_2Se_3} (M = Cu, Sr, Nb) while reported possible odd-parity pairing and zero-bias conductivity peaks in \ce{Sn_{1-x}In_xTe} point to the topological nature of this material \cite{Sasaki-SnInTe-TSC-data, Sato-SnInTe-topological, Schmidt-SnInTe-topological}.
Due to the immense interest in finding a true $bulk$ topological superconductor, as opposed to topological superconductivity achieved via proximity effect \cite{ProximityBeenakker} at the interface between a topological insulator and a conventional superconductor, it is vital to explore materials related to these compounds.
As several binary compounds in the Sn-In-Pb-Te family form, this family is an obvious one to explore.
The end members of (Sn,In,Pb)Te are well known.
SnTe has rocksalt structure ($Fm\bar{3}m$) and was the first material to be recognized as a topological crystalline insulator \cite{Hsieh-SnTe,Tanaka-SnTe}.
Below 100 K, SnTe undergoes a structural phase transition to a ferroelectric rhombohedral state (space group $R\bar{3}m$) \cite{Aggarwal-SnTe-R3m} and becomes superconducting at sub-kelvin temperatures \cite{Hein-SnTe-SC}.
PbTe is a well-known thermoelectric material \cite{Dughaish-PbTe, Lalonde-PbTe} which also has the $Fm\bar{3}m$ structure and is not believed to be superconducting in its pure form \cite{Kang-PbTeTl}.
InTe has a body-centered tetragonal lattice structure ($I4/mcm$) \cite{Hogg-InTe-structure} and superconducts at 2.18 K \cite{Tittmann-InTe}.

\begin{figure}[t]
    \includegraphics[width=1\columnwidth]{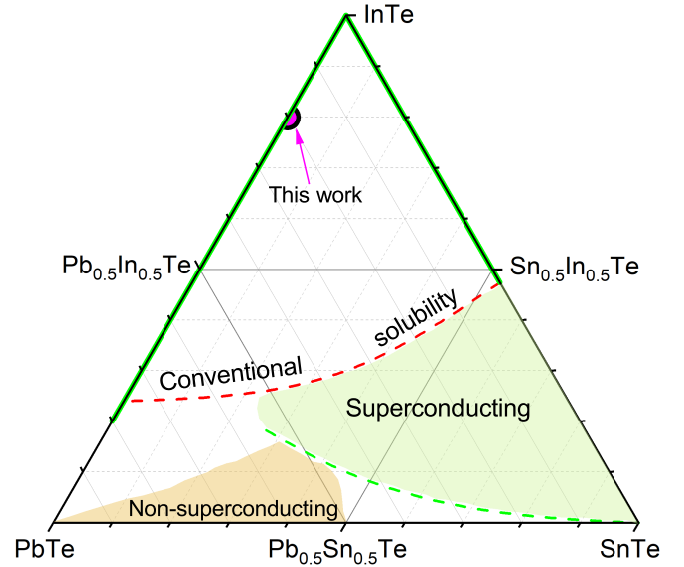}
    \caption{
Ternary phase diagram of the (Sn,In,Pb)Te system. 
Compositions above the red dashed line are above the solubility limit encountered in conventional synthesis; the green shaded area indicates the known superconducting region and the orange shaded area is non-superconducting.
HPHT synthesis has found superconductivity in \ce{Sn_{1-x}In_xTe} for all $x$ values and in \ce{In_{1-x}Pb_xTe} samples with $x < 0.8$ (solid green lines).
In this work we report on \ce{In_{0.8}Pb_{0.2}Te} (pink circle).
No \ce{(Pb_{1-x}Sn_x)_{1-y}In_yTe} obtained in HPHT synthesis has been reported.
}
    \label{figTernary}
\end{figure}

Upon doping SnTe with In, the topologically protected surface states near the Fermi energy are preserved \cite{Sato-SnInTe-topological, Schmidt-CMS-SnInTe} while the ferroelectric state is suppressed to $T = 0$ with 4\% In doping \cite{Erickson-SnInTe}.
There is a small window of $0 < x < 0.04$ in which \ce{Sn_{1-x}In_xTe} remains ferroelectric and superconducting \cite{Novak-SnInTe-disorder} with $T_c$ below 2 K, independent of $x$.
For doping levels above $\sim$4\%, $T_c$ increases linearly with doping to 4.5 K until the solubility limit is reached at $x \sim$50\% \cite{Zhong-SnInTe-growth}.
In contrast, PbTe is a trivial band insulator which undergoes a topological phase transition to a topological crystalline insulator phase upon Sn doping to 25\% \cite{Xu-NatComm-PbSnTe, Tanaka-PRB-PbSnTe}.
Doping PbTe with In through conventional synthesis techniques does not generate superconductivity up to the In solubility limit of $\sim$24\% \cite{Ravich-PbInTe,Rosenberg-PbInTe}.
Co-doping with Sn and In to form \ce{(Pb_{1-x}Sn_x)_{1-y}In_yTe} preserves the topological state for $x > 0.25$ and also generates potentially topological superconductivity \cite{Zhong-PbSnInTe-phase-diagram, Mikhailin-PbSnInTe, Denisov-PbSnInTe} in a limited In-doping window which depends on Sn concentration.
A ternary phase diagram following Zhong et al.~\cite{Zhong-PbSnInTe-phase-diagram} is shown in Fig.~\ref{figTernary}, where the region above the dashed red line is inaccessible by conventional synthesis techniques, the green shaded area indicates the known superconducting region, and the Pb-heavy orange shaded portion of the diagram is reported as nonsuperconducting.
Solubility limits can be bypassed by high pressure/high temperature (HPHT) synthesis \cite{Kobayashi-HPHT-SnInTe, Katsuno-HPHT-InPbTe}, allowing the entire range from SnTe to InTe and from PbTe to InTe to be explored.
Currently, only materials of the form \ce{In_{1-x}Pb_xTe} and \ce{Sn_{1-x}In_xTe} have been grown via HPHT synthesis, with superconductivity observed over a wide range of compositions (green lines in Fig.~\ref{figTernary}).
No \ce{(Pb_{1-x}Sn_x)_{1-y}In_yTe} HPHT synthesis has been reported yet.
When made via HPHT synthesis, all \ce{In_{1-x}Pb_xTe} compounds adopt the $Fm\bar{3}m$ FCC structure.
This phase seems to be metastable, reverting to the tetragonal structure when samples are kept at room temperature over extended time.
The end-line compound PbTe remains non-superconducting, whereas superconductivity in HPHT-synthesized InTe is observed with onset temperatures of 3 to 3.5 K (as compared to 2.18 K in the conventionally grown tetragonal material, see above).
While the FCC phase of InTe is believed to be topologically trivial, recent measurements and calculations \cite{Rajaji-InTe-TCI} suggest that doping InTe may in fact result in topologically nontrivial states as was the case with doping PbTe.
Reports show a full superconducting gap in conventionally synthesized \ce{(Pb_{0.5}Sn_{0.5})_{0.7}In_{0.3}Te} \cite{Du-PbSnTe-full-gap-STS} and also in \ce{Sn_{1-x}In_xTe} \cite{Smylie-SnInTe, Smylie-SnInTe-HPHT} for low and high In-doping.
No measurements on the low-Pb \ce{In_{1-x}Pb_xTe} phase have been reported yet.
It is imperative to measure the bulk properties of newly accessible doping regimes of InTe to look for evidence of an unconventional superconducting gap which may indicate topological behavior.

In this work, we report on magnetization and transport measurements at $^4$He temperatures and measurements of the temperature dependence of the London penetration depth $\Delta \lambda(T)$ down to $\sim$450 mK in superconducting \ce{In_{0.8}Pb_{0.2}Te} grown by HPHT synthesis (pink circle in Fig.~\ref{figTernary}).
The observed temperature dependence of $\Delta \lambda$ indicates a full superconducting gap, which eliminates one of two possible candidate topological superconducting states for this material.
We find the onset of superconductivity at $\sim4.75$ K and a lower and upper critical field of 9 mT and 2.12 T, respectively, for the bulk phase, with a corresponding estimate of 12.5 nm and 254 nm for the coherence length and the London penetration depth, respectively.
There is a small secondary phase likely consisting of a different doping level of \ce{In_{1-x}Pb_xTe} with an elevated onset $T_c$ (5.75 K) that has a substantially higher zero-temperature upper critical field estimated at 7.14 T.

\section{Experimental Methods}
Polycrystalline mm-scale ingots of \ce{In_{0.8}Pb_{0.2}}Te were grown following the method of Katsuno \cite{Katsuno-HPHT-InPbTe}.
Samples were kept in a freezer at approximately -20$^{\circ}$C to avoid a transition from the metastable NaCl-type structure to the tetragonal structure.
X-ray diffraction (XRD) measurement performed in a PANalytical X’pert Pro x-ray diffractometer with Cu $K_\alpha$ radiation on the ingot chosen for SQUID magnetization and Tunnel Diode Oscillator magnetometry (TDO) measurements confirmed the sample has the NaCl-type $Fm\bar{3}m$ structure with a lattice constant of 6.225 \AA, matching that of ingots measured immediately following synthesis.

Magnetization measurements were performed on an irregularly shaped ingot with a mass of 7.4 mg in a Quantum Design MPMS DC SQUID magnetometer at temperatures down to 1.8 K.
The sample was mounted inside a gelatin capsule held inside a clear plastic drinking straw.
The TDO technique \cite{ProzorovTDO} was used on a small piece (approximately 0.5 mm x 0.5 mm x 0.1 mm) cut from this ingot to measure the temperature dependence of the London penetration depth $\Delta \lambda(T) =\lambda(T)-\lambda_0$, where $\lambda_0$ is the zero-temperature value, down to $\sim$450 mK in an Oxford $^3$He cold-finger cryostat with a custom-built resonator \cite{SmylieTDO1} running at $\sim$14.5 MHz.

Due to the brittleness and small sizes of polycrystalline pieces, polishing a flat surface for 4-point contact geometry with silver paint contacts was not viable.
As a result, magneto-transport measurements were performed in a 2-point contact geometry using silver epoxy (Epotek H20E) and 50 $\mu$m gold wires, with the sample mounted on a custom-built probe in a cryostat capable of 1.6 K and 9 T.
Transport measurements were performed with a DC current of 50 $\mu$A.


\section{Results and Discussion}
\begin{figure}[t]
    \includegraphics[width=1\columnwidth]{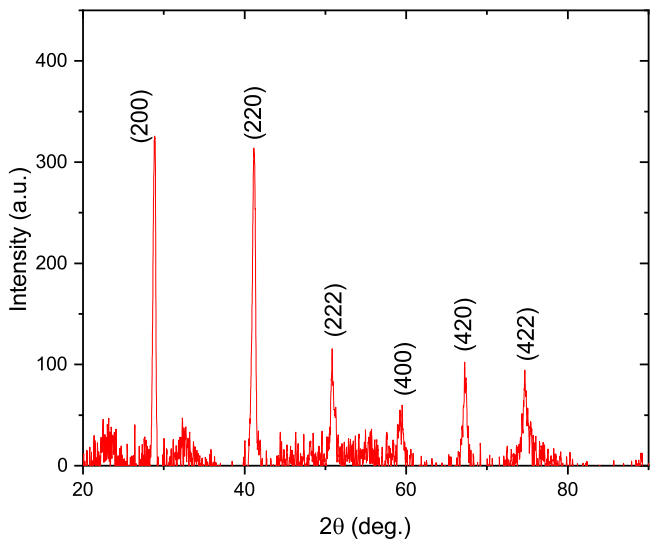}
    \caption{
Room temperature XRD pattern of a small piece of polycrystalline \ce{In_{0.8}Pb_{0.2}Te} following background subtraction.
All observable peaks above the noise level can be indexed with the NaCl-type $Fm\bar{3}m$ structure with a lattice constant 6.225 \AA.
}
\label{figXRD}
\end{figure}

Figure~\ref{figXRD} shows the room temperature XRD pattern of a small piece of polycrystalline \ce{In_{0.8}Pb_{0.2}Te} following background subtraction.
All peaks belong to the $Fm\bar{3}m$ space group, with no impurity peaks due to the tetragonal phase observed.
The lattice constant is found to be 6.225 Å, in agreement with previous reports for this doping level \cite{Katsuno-HPHT-InPbTe} despite the sample having been stored in a freezer for 18 months.
In the $Fm\bar{3}m$ space group, this yields an x-ray density of approximately 7.845 g/cm$^3$.

\begin{figure}[t]
    \includegraphics[width=1\columnwidth]{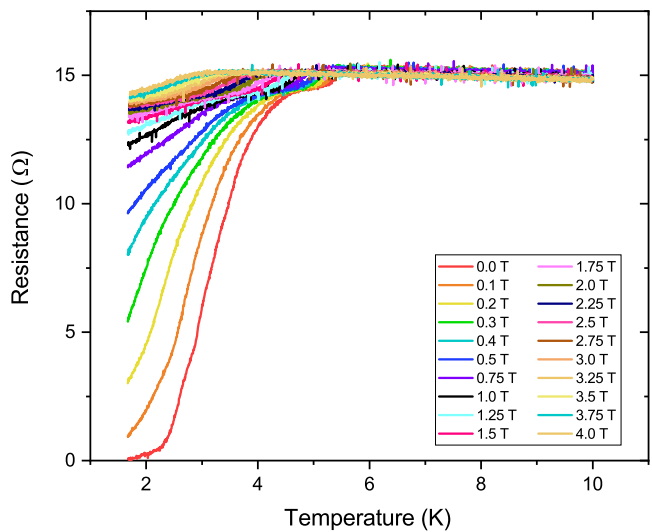}
    \caption{
Two-point resistance measurements as a function of temperature in various applied fields with a current of 50 $\mu$A after subtracting the residual contact resistance.
A broad transition with $T_{c,onset} \sim$4.5 K is observed, with a small secondary transition at $T_{c,onset}\sim$5.7 K.
Both transitions are suppressed with increasing field.
}
\label{figTransport}
\end{figure}

Low-temperature magneto-transport measurements are shown in Fig.~\ref{figTransport}.
A constant contact resistance stemming from the 2-point geometry has been subtracted.
Resistance measurements reveal a two-step transition with onsets of $T_{c,onset} \sim$5.7 K and 4.5 K, respectively.
Judging from the height of the resistive transitions, we attribute the transition at $\sim$4.5 K to the bulk $x = 0.2$ phase, and the transition at 5.7 K to a higher-$T_c$ impurity phase of \ce{In_{1-x}Pb_xTe} with a slightly different doping concentration; the exact relation between $T_c$ and $x$ is not fully understood \cite{Katsuno-HPHT-InPbTe}. Both transitions are suppressed in increasing magnetic fields.

\begin{figure}[t]
    \includegraphics[width=1\columnwidth]{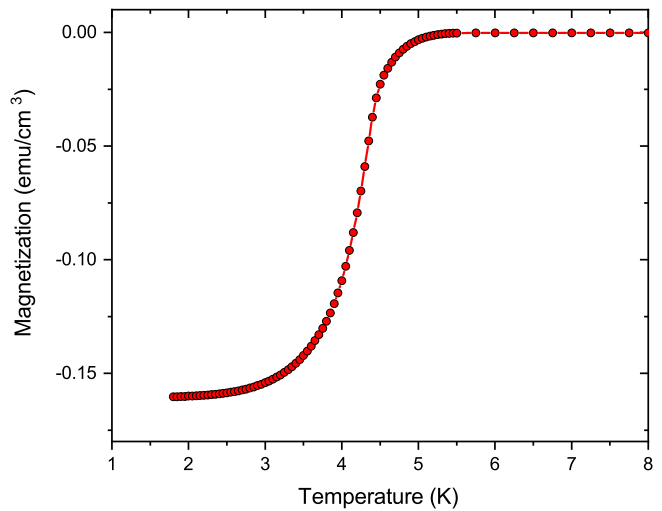}
    \caption{
Zero field-cooled magnetic susceptibility measurements in a field of 0.2 mT.
The transition has an onset at $\sim$5.25 K and strongly decreases at $\sim$4.75 K.
}
    \label{figMT2G}
\end{figure}

Zero field-cooled magnetization measurements using a DC SQUID magnetometer in an applied field of 0.2 mT are shown in Fig.~\ref{figMT2G}.
The sample is diamagnetic with an onset of $\sim$5.25 K.
The magnetization rapidly decreases below $\sim$4.75 K.

\begin{figure}[t]
    \includegraphics[width=1\columnwidth]{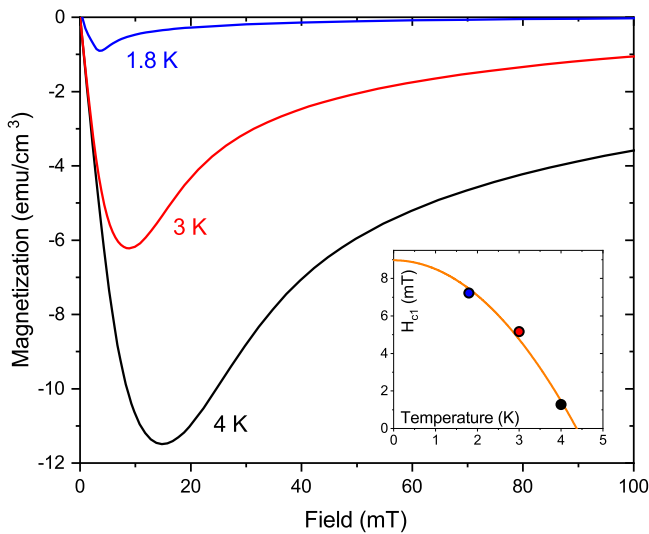}
    \caption{
Zero-field cooled magnetization vs applied field at multiple temperatures.
The penetration field, used to estimate $H_{c1}$, is taken as where the magnetization first deviates from linearity.
The inset shows $H_{c1}$ values after correcting for demagnetization effects and an extrapolation to the zero-temperature value, which we estimate at $\sim$9 mT.
}
\label{figMH}
\end{figure}

We determine the lower and upper critical fields $H_{c1}$ and $H_{c2}$ from field and temperature dependent magnetization measurements, respectively.
We deduce $H_{c1}$ from the value of the penetration field $H_p$ at which the magnetization starts deviating from the linear Meissner behavior at fixed temperature, with field sweeps at different temperatures as shown in Fig 5.
By comparing the slope of the $M$ vs $H$ measurements at low field and at 1.8 K to the ideal value of $-1/4\pi $ for a perfect diamagnet, we determine an effective demagnetization coefficient of approximately $N \approx 0.47$, yielding the $H_{c1} \approx H_p/(1-N)$ values shown in the inset of Fig.~\ref{figMH}.
This estimate neglects possible effects due to surface barriers, grain boundaries, and the geometric effects of the corners and sharp edges of the irregularly shaped piece of polycrystalline material.
The orange line in the inset of Fig.~\ref{figMH} represents a conventional parabolic temperature dependence $H_{c1}=H_{c1}(0)(1-(\frac{T}{T_c})^2)$ to the $H_{c1}$ datapoints, from which we extrapolate $\mu_0 H_{c1}$ at $T=0$ to be approximately 9 mT.

\begin{figure}[t]
    \includegraphics[width=1\columnwidth]{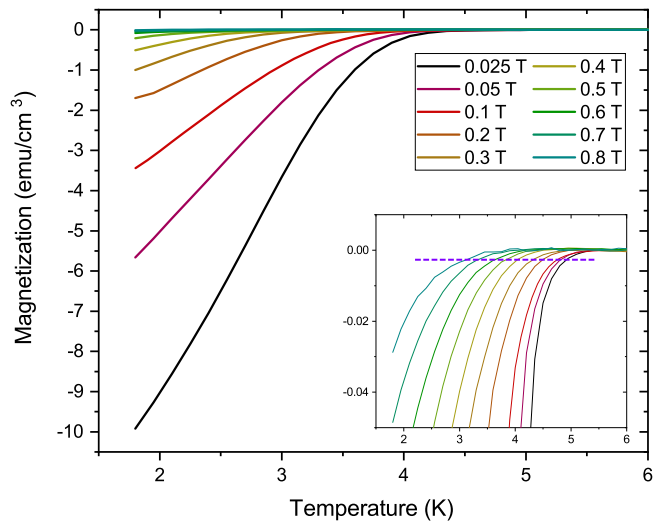}
    \caption{
Zero field-cooled magnetic susceptibility vs temperature for polycrystalline \ce{In_{1-x}Pb_xTe} with $x = 0.2$ as a function of temperature in different applied fields.
The inset shows the diamagnetic transition onsets after background subtraction, with the onset $T_c$ defined as where the transition crosses the dashed purple line.
}
\label{figMTH}
\end{figure}

Measurements of the temperature dependence of the magnetization in multiple applied fields via SQUID magnetometry are shown in Fig.~\ref{figMTH}.
As the field increases, two effects occur.
The superconducting transition shifts to progressively lower temperature, and the paramagnetic response of the Pb dopant becomes more pronounced; subtracting the normal state background yields the diamagnetic signals shown in the inset.
We define the onset $T_c$ as where each curve crosses the purple dashed line chosen close to zero but outside the noise level.
A different criterion will slightly shift the phase boundary but not affect its shape as discussed further below.

\begin{figure}[t]
    \includegraphics[width=1\columnwidth]{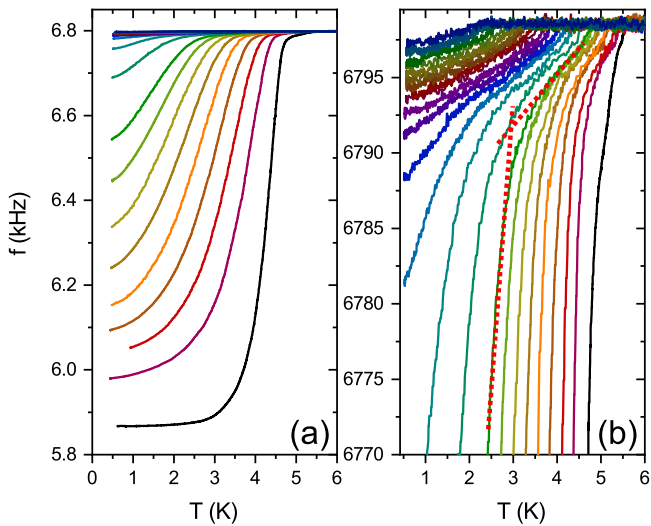}
    \caption{
(a) TDO resonator frequency shift vs temperature for a single piece of polycrystalline \ce{In_{1-x}Pb_xTe} with $x = 0.2$ in zero field (black) and in magnetic fields from 0.25 T to 5.5 T in 0.25 T steps.
The diamagnetic transition onset is shifted to lower temperatures as field is increased.
(b) Expanded view of the onset of the superconducting transitions; there is clearly a small second phase with higher $T_c$ which is suppressed more slowly with field.
The intersection of dashed red lines indicates the criterion used for determining $T_c$ of the lower-temperature phase.
}
\label{figTDOfTH}
\end{figure}

The shift of the TDO frequency with temperature and/or field is a measurement of the degree of screening of magnetic flux in the sample which is either due to superconductivity or the normal-state skin depth \cite{ProzorovTDO}.
In a TDO measurement, the transition into or out of the superconducting state is accompanied by a large shift in oscillator frequency, allowing mapping of the temperature dependence of the upper critical field.
TDO measurements on an irregular section cut from the ingot measured in the SQUID are shown in Fig.~\ref{figTDOfTH}(a) in applied magnetic field values of up to 5.5 T in 0.25 T steps.
In all measurements, the sample was field cooled from above $T_c$, then data were collected during a slow warming ramp through the transition and beyond.
As the resonant frequency of the TDO circuit is field-sensitive, the presented curves were all offset in frequency by constant amounts such that the TDO responses in the normal state aligned.
In zero field, there is a large diamagnetic transition with onset temperature of $\sim$4.69 K, which is continuously suppressed in field.
Zooming in near the onset of the transition [Fig.~\ref{figTDOfTH}(b)], there is clear evidence for a second phase with a higher onset temperature of  $\sim$5.75 K, which is suppressed in field at a slower rate than the bulk phase.
The onset temperature found here is comparable to the value observed in higher-doped \ce{In_{1-x}Pb_xTe} \cite{Katsuno-HPHT-InPbTe}, the most likely candidate for this minority phase.
The frequency shifts associated with each transition ($\sim$950 Hz for the bulk phase, $< 5$ Hz for the minority phase) allow for a rough estimation of the volume fraction indicating that this minority phase is a very small part of the bulk.

\begin{figure}[t]
    \includegraphics[width=1\columnwidth]{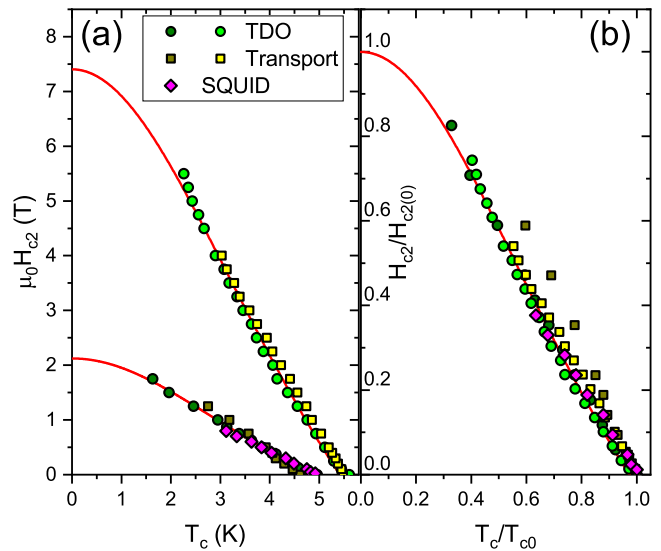}
    \caption{
(a) Superconducting phase diagram for polycrystalline \ce{In_{1-x}Pb_xTe} with $x = 0.2$ measured by multiple techniques.
SQUID measurements did not clearly distinguish the minority phase seen in TDO and magneto-transport data.
The red lines are a $\frac{1-T^2}{1+T^2}$ fit to the TDO data.
Extrapolated zero-temperature values for the upper critical field are 2.12 T and 7.4 T for the bulk phase and the minority phase, respectively.
(b) When transport and TDO curves are normalized to the extrapolated $T_{c0}$ and $H_{c2}(0)$ values, all data approximately falls on a universal curve, suggesting the minority phase is indeed the same material with a different doping level.
}
\label{figPhaseBoundaries}
\end{figure}

The TDO, SQUID and magneto-transport measurements can be used to determine a superconducting phase diagram which contains both the bulk phase and the higher-$T_c$ minority phase.
We define the transition temperature of the minority phase as the onset temperature where deviation from the temperature-independent normal-state behavior occurs.
Due to the higher transition temperature phase, the transition temperature in the bulk phase must be estimated as where the signal first deviates from linearity below the onset of the higher temperature phase; an example is shown in Fig.~\ref{figTDOfTH}(b) as the intersection of two dashed red lines superimposed on the $H$ = 1 T dataset.
The phase boundaries for both the bulk phase and the minority phase as derived from TDO measurements (circles), magneto-transport (squares), and SQUID measurements (diamonds) are shown in Fig.~\ref{figPhaseBoundaries}(a), with the minority phase having a higher onset $T_c$ and a steeper $dH_{c2}/dT$.
The minority phase was not clearly distinguishable in SQUID measurements; otherwise, there is very good agreement between all three techniques.
The $T_c$ criteria used here emphasize the onset of superconductivity and may overestimate $H_{c2}$ as compared to other techniques such as the resistive midpoints.
A phenomenological fit to the TDO data following $H_{c2} (T)= H_{c2}(0)(1-T^2)/(1+T^2)$ is shown in red in Fig.~\ref{figPhaseBoundaries}(a) to estimate the zero-temperature values of $H_{c2}$.
The fits extrapolate to an upper critical field at $T=0$ of $\mu_0 H_{c2} (0)=2.12$ T for the bulk phase and $\mu_0 H_{c2} (0)=7.14$ T for the minority phase; both are clearly lower than expected from the Pauli limit of $1.85*T_c$.
From the bulk phase value of $H_{c2} (0)$, we calculate the zero-temperature Ginzburg-Landau coherence length $\xi_0$ to be approximately 12.5 nm via the definition $\mu_{0} H_{c2}=\Phi_0/2\pi\xi^2 (0)$. 
Additionally, using the Ginzburg-Landau relation $ H_{c1}=\Phi_0/(4\pi\lambda^2)(ln[\lambda/\xi]+0.5)$ we estimate the zero-temperature London penetration depth $\lambda_0$ to be approximately 254 nm.
This value suggests a Ginzburg-Landau parameter $\kappa=\lambda/\xi$ of 20, identifying \ce{In_{1-x}Pb_xTe} as extreme type-II.
The two phase boundaries from all three techniques agree with each other.
In Fig.~\ref{figPhaseBoundaries}(b), we present the phase boundary data for both the bulk phase and the minority phase, with each dataset scaled by its own $T_c$ in zero field and by the extrapolated $H_{c2}$ at $T=0$ values.
With this scaling, all datasets fall on a universal line.
We take this as evidence that the minority phase is indeed just a higher doping level of the bulk phase.
There are no other known candidates consisting of Pb, In, or Te in this $T_c$ range or with such high critical fields.

Low-temperature penetration depth measurements were carried out via the TDO technique in the temperature range of 0.45 K to 10 K.
In the TDO technique, the frequency shift $\Delta f$ of the resonator is proportional to the change of the penetration depth \cite{ProzorovTDO, Carrington-TDO-1999, Prozorov-Giannetta-APL-2000}:
\begin{equation}
\Delta f(T) = G\Delta \lambda(T)
\label{eq:F(T)}
\end{equation}
where the geometrical factor G depends on the geometry of the resonator coil as well as the sample volume and shape.
The small amplitude of the AC field in the coil ($\sim$2 $\mu$T) ensures that the sample remains fully in the Meissner state below $T_c$.
The low-temperature variation of the London penetration depth $\Delta\lambda(T)=\lambda(T)-\lambda_0$ can provide information on the superconducting gap structure \cite{ProzorovTDO}. 
In the limit of low temperature, conventional BCS theory for an isotropic s-wave superconductor yields an exponential variation of $\Delta\lambda(T)$:
\begin{equation}
\frac{\Delta\lambda(T)}{\lambda_0} \approx \sqrt{\frac{\pi\Delta_0}{2T}}\exp \left (\frac{-\Delta_0}{T}\right)
\end{equation}
where $\Delta_0$ and $\lambda_0$ are the zero-temperature values of the superconducting energy gap and the penetration depth.
In nodal superconductors, however, a stronger temperature dependence is observed due to the enhanced thermal excitation of quasiparticles near the gap nodes, where the gap amplitude decreases.
This results in a power-law variation in the penetration depth, $\Delta\lambda \sim T^n$, where the exponent n depends on the nature of the nodes and the degree of electron scattering; line nodes in the energy gap will generate a $T$-linear response, whereas point nodes will generate a $T^2$ response.
Behavior with $n\geq3$ is generally considered indistinguishable from exponential and is taken as evidence of a full superconducting gap.

\begin{figure}
    \includegraphics[width=1\columnwidth]{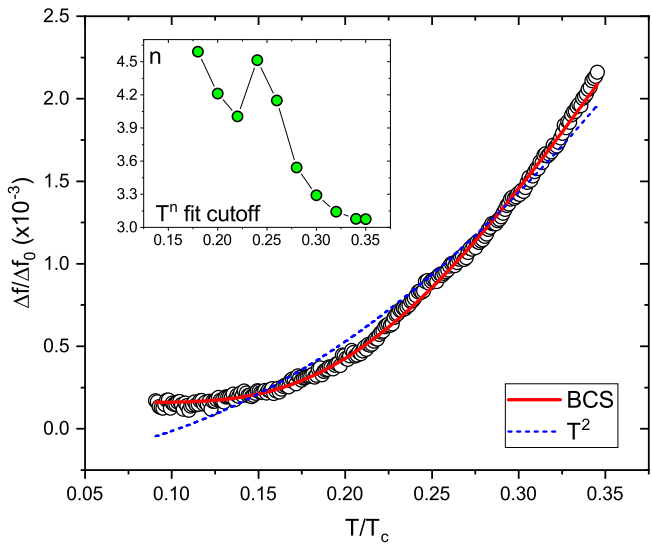}
    \caption{
Normalized low-temperature frequency shift $\Delta f(T)$ for polycrystalline \ce{In_{1-x}Pb_xTe} with $x = 0.2$.
The BCS-like fit (red) well describes the data; a nodal $T^2$ fit (blue, dashed) poorly describes the data.
(inset) Exponent of a $T^n$ fit to the data vs reduced temperature cutoff for the fit; fits with $n\geq3$ are generally taken as indistinguishable from exponential.
}
\label{figLambdaT}
\end{figure}

The evolution of the normalized low-temperature TDO frequency shift of \ce{In_{1-x}Pb_xTe} with $x = 0.2$ is shown in Fig.~\ref{figLambdaT}.
Instrument noise is on the order of 0.25 Hz, and with averaging slow temperature sweeps back and forth between 0.42 K and 1.6 K, a noise level of $< 0.05$ Hz is achievable, with a full superconducting transition frequency shift of $\sim$950 Hz.
The variation of the frequency shift can be well described (red line) by an exponential dependence with a BCS-like gap value across a wide range of temperature.
The observed gap ratio of $\Delta_0/T_c=1.07$ is smaller than the standard BCS value $\Delta_0/T_c= 1.76$ for an isotropic single-gap s-wave superconductor.
A possible cause could be a sizable anisotropy of the gap \cite{AnisotropicGap1, AnisotropicGap2, AnisotropicGap3}.
Furthermore, low-energy features found in inelastic neutron scattering measurements \cite{Ran-PbSnInTe, Sapkota-PbSnInTe} are not consistent with conventional BCS-like behavior, and calorimetry measurements \cite{Katsuno-HPHT-InPbTe} across the (Pb,In)Te family shows an abrupt dip in the Sommerfeld coefficient in the doping window where $T_c$ is approximately constant, suggesting superconductivity in this very low carrier density family of compounds may not be entirely BCS-like.
A $T^2$ fit, characteristic of nodal behavior (dashed, blue), poorly describes the data.
The inset shows the exponent in a fit of the form $AT^n$ with the exponent $n$ a free parameter, while the fit is calculated from the minimum temperature to various cutoff values of reduced temperature $t=T/T_c$.
Over the entire low-temperature range, the fitted exponent $n$ is at least 3, and increases quickly as the upper temperature cutoff for the fit decreases.
Such a fit is essentially indistinguishable from the exponential, BCS-like response shown in red in the main panel.
The presence of the higher-temperature phase does not affect the fit due to its small relative volume fraction as seen in the TDO data.

\section{Conclusion}
In summary, we have investigated the superconducting gap symmetry and critical fields of polycrystalline heavily In-doped PbTe, with doping well beyond the previously known In saturation limit.
This doping regime, only available via high-pressure synthesis techniques, shows a higher $T_c$ while maintaining an FCC crystal structure.
A small minority phase appears with higher $T_c$ than the bulk and is most probably associated with higher Pb-doped material due to inhomogenous doping in the growth; both phases have the same normalized magnetic phase boundaries down to sub-K temperatures.
The superconducting gap structure is full and likely anisotropic, and the material is an extreme type-II superconductor with a $\kappa$ of approximately 20.
The zero-temperature upper critical field in the $\sim$5.7 K phase is 7.14 T, and is quite a bit higher compared with another report \cite{Katsuno-HPHT-InPbTe} on HPHT-synthesized \ce{In_{1-x}Pb_{x}Te} which found $\mu_0 H_{c2}=3.1$ T for $T_c$ of $\sim$5.3 K.
Katsuno reports a nonlinear decrease in the Sommerfeld coefficient $\gamma$ with increasing Pb doping and increasing $T_c$, with $\gamma$ sharply decreasing near max $T_c$.
This is contrary to BCS expectations of increasing $\gamma$ corresponding to increasing density of states and thus to higher $T_c$ values.
Further investigation across the doping diagram is necessary to understand the origin of this behavior.
Regarding any possible topological nature of the superconducting state, extending the symmetry arguments applied to the isostructural superconducting topological insulator \ce{Sn_{1-x}In_xTe} \cite{Sasaki-SnInTe-TSC-data} suggests the superconducting gap in highly doped \ce{In_{1-x}Pb_{x}Te} is either the conventional, topologically trivial $A_{1g}$ state, or the odd-parity, topologically nontrivial $A_{1u}$ state.
We eliminate the odd-parity, topologically nontrivial nodal $A_{2u}$ state as a possibility.
Further synthesis and study of the newly available region of the ternary phase diagram (Fig.~\ref{figTernary}) via HPHT synthesis, especially of the \ce{(Pb_{1-x}Sn_x)_{1-y}In_yTe} formulations, should be pursued to look for superconductivity in topologically nontrivial compounds.

\section*{Acknowledgements}
TDO and magnetization measurements at Argonne were supported by the US Department of Energy, Office of Science, Basic Energy Sciences, Materials Sciences and Engineering Division.
Crystal synthesis was supported by Grant-in-Aid from the Ministry of Education, Culture, Sports, Science, and Technology (MEXT) under Grants No. 18K03540 and No. 19H01852 and No. 20K05272.

\bibliography{InPbTe.bib}

 \newcommand{\noop}[1]{}
\begin{thebibliography}{55}%
\makeatletter
\providecommand \@ifxundefined [1]{%
 \@ifx{#1\undefined}
}%
\providecommand \@ifnum [1]{%
 \ifnum #1\expandafter \@firstoftwo
 \else \expandafter \@secondoftwo
 \fi
}%
\providecommand \@ifx [1]{%
 \ifx #1\expandafter \@firstoftwo
 \else \expandafter \@secondoftwo
 \fi
}%
\providecommand \natexlab [1]{#1}%
\providecommand \enquote  [1]{``#1''}%
\providecommand \bibnamefont  [1]{#1}%
\providecommand \bibfnamefont [1]{#1}%
\providecommand \citenamefont [1]{#1}%
\providecommand \href@noop [0]{\@secondoftwo}%
\providecommand \href [0]{\begingroup \@sanitize@url \@href}%
\providecommand \@href[1]{\@@startlink{#1}\@@href}%
\providecommand \@@href[1]{\endgroup#1\@@endlink}%
\providecommand \@sanitize@url [0]{\catcode `\\12\catcode `\$12\catcode
  `\&12\catcode `\#12\catcode `\^12\catcode `\_12\catcode `\%12\relax}%
\providecommand \@@startlink[1]{}%
\providecommand \@@endlink[0]{}%
\providecommand \url  [0]{\begingroup\@sanitize@url \@url }%
\providecommand \@url [1]{\endgroup\@href {#1}{\urlprefix }}%
\providecommand \urlprefix  [0]{URL }%
\providecommand \Eprint [0]{\href }%
\providecommand \doibase [0]{https://doi.org/}%
\providecommand \selectlanguage [0]{\@gobble}%
\providecommand \bibinfo  [0]{\@secondoftwo}%
\providecommand \bibfield  [0]{\@secondoftwo}%
\providecommand \translation [1]{[#1]}%
\providecommand \BibitemOpen [0]{}%
\providecommand \bibitemStop [0]{}%
\providecommand \bibitemNoStop [0]{.\EOS\space}%
\providecommand \EOS [0]{\spacefactor3000\relax}%
\providecommand \BibitemShut  [1]{\csname bibitem#1\endcsname}%
\let\auto@bib@innerbib\@empty
\bibitem [{\citenamefont {Fu}\ \emph {et~al.}(2007)\citenamefont {Fu},
  \citenamefont {Kane},\ and\ \citenamefont {Mele}}]{Fu-TI-prediction}%
  \BibitemOpen
  \bibfield  {author} {\bibinfo {author} {\bibfnamefont {L.}~\bibnamefont
  {Fu}}, \bibinfo {author} {\bibfnamefont {C.~L.}\ \bibnamefont {Kane}},\ and\
  \bibinfo {author} {\bibfnamefont {E.~J.}\ \bibnamefont {Mele}},\ }\href
  {https://doi.org/10.1103/PhysRevLett.98.106803} {\bibfield  {journal}
  {\bibinfo  {journal} {Phys. Rev. Lett.}\ }\textbf {\bibinfo {volume} {98}},\
  \bibinfo {pages} {106803} (\bibinfo {year} {2007})}\BibitemShut {NoStop}%
\bibitem [{\citenamefont {{Zhang}}\ \emph {et~al.}(2009)\citenamefont
  {{Zhang}}, \citenamefont {{Liu}}, \citenamefont {{Qi}}, \citenamefont
  {{Dai}}, \citenamefont {{Fang}},\ and\ \citenamefont
  {{Zhang}}}]{Zhang-Bi2Se3-predict}%
  \BibitemOpen
  \bibfield  {author} {\bibinfo {author} {\bibfnamefont {H.}~\bibnamefont
  {{Zhang}}}, \bibinfo {author} {\bibfnamefont {C.-X.}\ \bibnamefont {{Liu}}},
  \bibinfo {author} {\bibfnamefont {X.-L.}\ \bibnamefont {{Qi}}}, \bibinfo
  {author} {\bibfnamefont {X.}~\bibnamefont {{Dai}}}, \bibinfo {author}
  {\bibfnamefont {Z.}~\bibnamefont {{Fang}}},\ and\ \bibinfo {author}
  {\bibfnamefont {S.-C.}\ \bibnamefont {{Zhang}}},\ }\href
  {https://doi.org/10.1038/nphys1270} {\bibfield  {journal} {\bibinfo
  {journal} {Nat. Phys.}\ }\textbf {\bibinfo {volume} {5}},\ \bibinfo {pages}
  {438} (\bibinfo {year} {2009})}\BibitemShut {NoStop}%
\bibitem [{\citenamefont {Xia}\ \emph {et~al.}(2009)\citenamefont {Xia},
  \citenamefont {Qian}, \citenamefont {Hsieh}, \citenamefont {Wray},
  \citenamefont {Pal}, \citenamefont {Lin}, \citenamefont {Bansil},
  \citenamefont {Grauer}, \citenamefont {Hor}, \citenamefont {Cava},\ and\
  \citenamefont {Hasan}}]{XiaBi2Se3-discovery}%
  \BibitemOpen
  \bibfield  {author} {\bibinfo {author} {\bibfnamefont {Y.}~\bibnamefont
  {Xia}}, \bibinfo {author} {\bibfnamefont {D.}~\bibnamefont {Qian}}, \bibinfo
  {author} {\bibfnamefont {D.}~\bibnamefont {Hsieh}}, \bibinfo {author}
  {\bibfnamefont {L.}~\bibnamefont {Wray}}, \bibinfo {author} {\bibfnamefont
  {A.}~\bibnamefont {Pal}}, \bibinfo {author} {\bibfnamefont {H.}~\bibnamefont
  {Lin}}, \bibinfo {author} {\bibfnamefont {A.}~\bibnamefont {Bansil}},
  \bibinfo {author} {\bibfnamefont {D.}~\bibnamefont {Grauer}}, \bibinfo
  {author} {\bibfnamefont {Y.}~\bibnamefont {Hor}}, \bibinfo {author}
  {\bibfnamefont {R.}~\bibnamefont {Cava}},\ and\ \bibinfo {author}
  {\bibfnamefont {M.}~\bibnamefont {Hasan}},\ }\href
  {https://doi.org/10.1038/nphys1274} {\bibfield  {journal} {\bibinfo
  {journal} {Nature Physics}\ }\textbf {\bibinfo {volume} {5}},\ \bibinfo
  {pages} {398} (\bibinfo {year} {2009})}\BibitemShut {NoStop}%
\bibitem [{\citenamefont {Hsieh}\ \emph {et~al.}(2009)\citenamefont {Hsieh},
  \citenamefont {Xia}, \citenamefont {Qian}, \citenamefont {Wray},
  \citenamefont {Dil}, \citenamefont {Meier}, \citenamefont {Osterwalder},
  \citenamefont {Patthey}, \citenamefont {Checkelsky}, \citenamefont {Ong},
  \citenamefont {Fedorov}, \citenamefont {Lin}, \citenamefont {Bansil},
  \citenamefont {Grauer}, \citenamefont {Hor}, \citenamefont {Cava},\ and\
  \citenamefont {Hasan}}]{HsiehBi2Se3-discovery}%
  \BibitemOpen
  \bibfield  {author} {\bibinfo {author} {\bibfnamefont {D.}~\bibnamefont
  {Hsieh}}, \bibinfo {author} {\bibfnamefont {Y.}~\bibnamefont {Xia}}, \bibinfo
  {author} {\bibfnamefont {D.}~\bibnamefont {Qian}}, \bibinfo {author}
  {\bibfnamefont {L.}~\bibnamefont {Wray}}, \bibinfo {author} {\bibfnamefont
  {J.}~\bibnamefont {Dil}}, \bibinfo {author} {\bibfnamefont {F.}~\bibnamefont
  {Meier}}, \bibinfo {author} {\bibfnamefont {J.}~\bibnamefont {Osterwalder}},
  \bibinfo {author} {\bibfnamefont {L.}~\bibnamefont {Patthey}}, \bibinfo
  {author} {\bibfnamefont {J.}~\bibnamefont {Checkelsky}}, \bibinfo {author}
  {\bibfnamefont {N.}~\bibnamefont {Ong}}, \bibinfo {author} {\bibfnamefont
  {A.}~\bibnamefont {Fedorov}}, \bibinfo {author} {\bibfnamefont
  {H.}~\bibnamefont {Lin}}, \bibinfo {author} {\bibfnamefont {A.}~\bibnamefont
  {Bansil}}, \bibinfo {author} {\bibfnamefont {D.}~\bibnamefont {Grauer}},
  \bibinfo {author} {\bibfnamefont {Y.}~\bibnamefont {Hor}}, \bibinfo {author}
  {\bibfnamefont {R.}~\bibnamefont {Cava}},\ and\ \bibinfo {author}
  {\bibfnamefont {M.}~\bibnamefont {Hasan}},\ }\href
  {https://doi.org/10.1038/nature08234} {\bibfield  {journal} {\bibinfo
  {journal} {Nature}\ }\textbf {\bibinfo {volume} {460}},\ \bibinfo {pages}
  {1101} (\bibinfo {year} {2009})}\BibitemShut {NoStop}%
\bibitem [{\citenamefont {Mazumder}\ and\ \citenamefont
  {Shirage}(2021)}]{Mazumder-Bi2Se3}%
  \BibitemOpen
  \bibfield  {author} {\bibinfo {author} {\bibfnamefont {K.}~\bibnamefont
  {Mazumder}}\ and\ \bibinfo {author} {\bibfnamefont {P.~M.}\ \bibnamefont
  {Shirage}},\ }\href
  {https://doi.org/https://doi.org/10.1016/j.jallcom.2021.161492} {\bibfield
  {journal} {\bibinfo  {journal} {Journal of Alloys and Compounds}\ }\textbf
  {\bibinfo {volume} {888}},\ \bibinfo {pages} {161492} (\bibinfo {year}
  {2021})}\BibitemShut {NoStop}%
\bibitem [{\citenamefont {Hsieh}\ \emph {et~al.}(2012)\citenamefont {Hsieh},
  \citenamefont {Lin}, \citenamefont {Liu}, \citenamefont {Duan}, \citenamefont
  {Bansil},\ and\ \citenamefont {Fu}}]{Hsieh-SnTe}%
  \BibitemOpen
  \bibfield  {author} {\bibinfo {author} {\bibfnamefont {T.~H.}\ \bibnamefont
  {Hsieh}}, \bibinfo {author} {\bibfnamefont {H.}~\bibnamefont {Lin}}, \bibinfo
  {author} {\bibfnamefont {J.}~\bibnamefont {Liu}}, \bibinfo {author}
  {\bibfnamefont {W.}~\bibnamefont {Duan}}, \bibinfo {author} {\bibfnamefont
  {A.}~\bibnamefont {Bansil}},\ and\ \bibinfo {author} {\bibfnamefont
  {L.}~\bibnamefont {Fu}},\ }\href@noop {} {\bibfield  {journal} {\bibinfo
  {journal} {Nature Communications}\ }\textbf {\bibinfo {volume} {3}} (\bibinfo
  {year} {2012})}\BibitemShut {NoStop}%
\bibitem [{\citenamefont {Tanaka}\ \emph {et~al.}(2012)\citenamefont {Tanaka},
  \citenamefont {Ren}, \citenamefont {Sato}, \citenamefont {Nakayama},
  \citenamefont {Souma}, \citenamefont {Takahashi}, \citenamefont {Segawa},\
  and\ \citenamefont {Ando}}]{Tanaka-SnTe}%
  \BibitemOpen
  \bibfield  {author} {\bibinfo {author} {\bibfnamefont {Y.}~\bibnamefont
  {Tanaka}}, \bibinfo {author} {\bibfnamefont {Z.}~\bibnamefont {Ren}},
  \bibinfo {author} {\bibfnamefont {T.}~\bibnamefont {Sato}}, \bibinfo {author}
  {\bibfnamefont {K.}~\bibnamefont {Nakayama}}, \bibinfo {author}
  {\bibfnamefont {S.}~\bibnamefont {Souma}}, \bibinfo {author} {\bibfnamefont
  {T.}~\bibnamefont {Takahashi}}, \bibinfo {author} {\bibfnamefont
  {K.}~\bibnamefont {Segawa}},\ and\ \bibinfo {author} {\bibfnamefont
  {Y.}~\bibnamefont {Ando}},\ }\href {https://doi.org/10.1038/nphys2442}
  {\bibfield  {journal} {\bibinfo  {journal} {Nature Physics}\ }\textbf
  {\bibinfo {volume} {8}},\ \bibinfo {pages} {800} (\bibinfo {year}
  {2012})}\BibitemShut {NoStop}%
\bibitem [{\citenamefont {Hor}\ \emph {et~al.}(2010)\citenamefont {Hor},
  \citenamefont {Williams}, \citenamefont {Checkelsky}, \citenamefont
  {Roushan}, \citenamefont {Seo}, \citenamefont {Xu}, \citenamefont
  {Zandbergen}, \citenamefont {Yazdani}, \citenamefont {Ong},\ and\
  \citenamefont {Cava}}]{Hor-CBS}%
  \BibitemOpen
  \bibfield  {author} {\bibinfo {author} {\bibfnamefont {Y.~S.}\ \bibnamefont
  {Hor}}, \bibinfo {author} {\bibfnamefont {A.~J.}\ \bibnamefont {Williams}},
  \bibinfo {author} {\bibfnamefont {J.~G.}\ \bibnamefont {Checkelsky}},
  \bibinfo {author} {\bibfnamefont {P.}~\bibnamefont {Roushan}}, \bibinfo
  {author} {\bibfnamefont {J.}~\bibnamefont {Seo}}, \bibinfo {author}
  {\bibfnamefont {Q.}~\bibnamefont {Xu}}, \bibinfo {author} {\bibfnamefont
  {H.~W.}\ \bibnamefont {Zandbergen}}, \bibinfo {author} {\bibfnamefont
  {A.}~\bibnamefont {Yazdani}}, \bibinfo {author} {\bibfnamefont {N.~P.}\
  \bibnamefont {Ong}},\ and\ \bibinfo {author} {\bibfnamefont {R.~J.}\
  \bibnamefont {Cava}},\ }\href
  {https://doi.org/10.1103/PhysRevLett.104.057001} {\bibfield  {journal}
  {\bibinfo  {journal} {Phys. Rev. Lett.}\ }\textbf {\bibinfo {volume} {104}},\
  \bibinfo {pages} {057001} (\bibinfo {year} {2010})}\BibitemShut {NoStop}%
\bibitem [{\citenamefont {Qiu}\ \emph {et~al.}(2015)\citenamefont {Qiu},
  \citenamefont {Sanders}, \citenamefont {Dai}, \citenamefont {Medvedeva},
  \citenamefont {Wu}, \citenamefont {Ghaemi}, \citenamefont {Vojta},\ and\
  \citenamefont {Hor}}]{Qiu-NBS}%
  \BibitemOpen
  \bibfield  {author} {\bibinfo {author} {\bibfnamefont {Y.}~\bibnamefont
  {Qiu}}, \bibinfo {author} {\bibfnamefont {K.~N.}\ \bibnamefont {Sanders}},
  \bibinfo {author} {\bibfnamefont {J.}~\bibnamefont {Dai}}, \bibinfo {author}
  {\bibfnamefont {J.~E.}\ \bibnamefont {Medvedeva}}, \bibinfo {author}
  {\bibfnamefont {W.}~\bibnamefont {Wu}}, \bibinfo {author} {\bibfnamefont
  {P.}~\bibnamefont {Ghaemi}}, \bibinfo {author} {\bibfnamefont
  {T.}~\bibnamefont {Vojta}},\ and\ \bibinfo {author} {\bibfnamefont {Y.~S.}\
  \bibnamefont {Hor}},\ }\Eprint {https://arxiv.org/abs/1512.03519}
  {arXiv:1512.03519}  (\bibinfo {year} {2015})\BibitemShut {NoStop}%
\bibitem [{\citenamefont {Liu}\ \emph {et~al.}(2015)\citenamefont {Liu},
  \citenamefont {Yao}, \citenamefont {Shao}, \citenamefont {Zuo}, \citenamefont
  {Pi}, \citenamefont {Tan}, \citenamefont {Zhang},\ and\ \citenamefont
  {Zhang}}]{Liu-SBS}%
  \BibitemOpen
  \bibfield  {author} {\bibinfo {author} {\bibfnamefont {Z.}~\bibnamefont
  {Liu}}, \bibinfo {author} {\bibfnamefont {X.}~\bibnamefont {Yao}}, \bibinfo
  {author} {\bibfnamefont {J.}~\bibnamefont {Shao}}, \bibinfo {author}
  {\bibfnamefont {M.}~\bibnamefont {Zuo}}, \bibinfo {author} {\bibfnamefont
  {L.}~\bibnamefont {Pi}}, \bibinfo {author} {\bibfnamefont {S.}~\bibnamefont
  {Tan}}, \bibinfo {author} {\bibfnamefont {C.}~\bibnamefont {Zhang}},\ and\
  \bibinfo {author} {\bibfnamefont {Y.}~\bibnamefont {Zhang}},\ }\href@noop {}
  {\bibfield  {journal} {\bibinfo  {journal} {Journal of the American Chemical
  Society}\ }\textbf {\bibinfo {volume} {137}},\ \bibinfo {pages} {10512}
  (\bibinfo {year} {2015})}\BibitemShut {NoStop}%
\bibitem [{\citenamefont {Zhong}\ \emph {et~al.}(2014)\citenamefont {Zhong},
  \citenamefont {Schneeloch}, \citenamefont {Liu}, \citenamefont {Camino},
  \citenamefont {Tranquada},\ and\ \citenamefont
  {Gu}}]{Zhong-PbSnInTe-superconductivity}%
  \BibitemOpen
  \bibfield  {author} {\bibinfo {author} {\bibfnamefont {R.~D.}\ \bibnamefont
  {Zhong}}, \bibinfo {author} {\bibfnamefont {J.~A.}\ \bibnamefont
  {Schneeloch}}, \bibinfo {author} {\bibfnamefont {T.~S.}\ \bibnamefont {Liu}},
  \bibinfo {author} {\bibfnamefont {F.~E.}\ \bibnamefont {Camino}}, \bibinfo
  {author} {\bibfnamefont {J.~M.}\ \bibnamefont {Tranquada}},\ and\ \bibinfo
  {author} {\bibfnamefont {G.~D.}\ \bibnamefont {Gu}},\ }\href
  {https://doi.org/10.1103/PhysRevB.90.020505} {\bibfield  {journal} {\bibinfo
  {journal} {Phys. Rev. B}\ }\textbf {\bibinfo {volume} {90}},\ \bibinfo
  {pages} {020505(R)} (\bibinfo {year} {2014})}\BibitemShut {NoStop}%
\bibitem [{\citenamefont {Erickson}\ \emph {et~al.}(2009)\citenamefont
  {Erickson}, \citenamefont {Chu}, \citenamefont {Toney}, \citenamefont
  {Geballe},\ and\ \citenamefont {Fisher}}]{Erickson-SnInTe}%
  \BibitemOpen
  \bibfield  {author} {\bibinfo {author} {\bibfnamefont {A.~S.}\ \bibnamefont
  {Erickson}}, \bibinfo {author} {\bibfnamefont {J.-H.}\ \bibnamefont {Chu}},
  \bibinfo {author} {\bibfnamefont {M.~F.}\ \bibnamefont {Toney}}, \bibinfo
  {author} {\bibfnamefont {T.~H.}\ \bibnamefont {Geballe}},\ and\ \bibinfo
  {author} {\bibfnamefont {I.~R.}\ \bibnamefont {Fisher}},\ }\href
  {https://doi.org/10.1103/PhysRevB.79.024520} {\bibfield  {journal} {\bibinfo
  {journal} {Phys. Rev. B}\ }\textbf {\bibinfo {volume} {79}},\ \bibinfo
  {pages} {024520} (\bibinfo {year} {2009})}\BibitemShut {NoStop}%
\bibitem [{\citenamefont {Mizuguchi}\ and\ \citenamefont
  {Miura}(2016)}]{Mizuguchi-SnAgTe}%
  \BibitemOpen
  \bibfield  {author} {\bibinfo {author} {\bibfnamefont {Y.}~\bibnamefont
  {Mizuguchi}}\ and\ \bibinfo {author} {\bibfnamefont {O.}~\bibnamefont
  {Miura}},\ }\href {https://doi.org/10.7566/JPSJ.85.053702} {\bibfield
  {journal} {\bibinfo  {journal} {Journal of the Physical Society of Japan}\
  }\textbf {\bibinfo {volume} {85}},\ \bibinfo {pages} {053702} (\bibinfo
  {year} {2016})}\BibitemShut {NoStop}%
\bibitem [{\citenamefont {Polley}\ \emph {et~al.}(2016)\citenamefont {Polley},
  \citenamefont {Jovic}, \citenamefont {Su}, \citenamefont {Saghir},
  \citenamefont {Newby}, \citenamefont {Kowalski}, \citenamefont {Jakiela},
  \citenamefont {Barcz}, \citenamefont {Guziewicz}, \citenamefont
  {Balasubramanian}, \citenamefont {Balakrishnan}, \citenamefont {Laverock},\
  and\ \citenamefont {Smith}}]{Polley-SnInTe-ARPES}%
  \BibitemOpen
  \bibfield  {author} {\bibinfo {author} {\bibfnamefont {C.~M.}\ \bibnamefont
  {Polley}}, \bibinfo {author} {\bibfnamefont {V.}~\bibnamefont {Jovic}},
  \bibinfo {author} {\bibfnamefont {T.-Y.}\ \bibnamefont {Su}}, \bibinfo
  {author} {\bibfnamefont {M.}~\bibnamefont {Saghir}}, \bibinfo {author}
  {\bibfnamefont {D.}~\bibnamefont {Newby}}, \bibinfo {author} {\bibfnamefont
  {B.~J.}\ \bibnamefont {Kowalski}}, \bibinfo {author} {\bibfnamefont
  {R.}~\bibnamefont {Jakiela}}, \bibinfo {author} {\bibfnamefont
  {A.}~\bibnamefont {Barcz}}, \bibinfo {author} {\bibfnamefont
  {M.}~\bibnamefont {Guziewicz}}, \bibinfo {author} {\bibfnamefont
  {T.}~\bibnamefont {Balasubramanian}}, \bibinfo {author} {\bibfnamefont
  {G.}~\bibnamefont {Balakrishnan}}, \bibinfo {author} {\bibfnamefont
  {J.}~\bibnamefont {Laverock}},\ and\ \bibinfo {author} {\bibfnamefont
  {K.~E.}\ \bibnamefont {Smith}},\ }\href
  {https://doi.org/10.1103/PhysRevB.93.075132} {\bibfield  {journal} {\bibinfo
  {journal} {Phys. Rev. B}\ }\textbf {\bibinfo {volume} {93}},\ \bibinfo
  {pages} {075132} (\bibinfo {year} {2016})}\BibitemShut {NoStop}%
\bibitem [{\citenamefont {Sato}\ \emph {et~al.}(2013)\citenamefont {Sato},
  \citenamefont {Tanaka}, \citenamefont {Nakayama}, \citenamefont {Souma},
  \citenamefont {Takahashi}, \citenamefont {Sasaki}, \citenamefont {Ren},
  \citenamefont {Taskin}, \citenamefont {Segawa},\ and\ \citenamefont
  {Ando}}]{Sato-SnInTe-topological}%
  \BibitemOpen
  \bibfield  {author} {\bibinfo {author} {\bibfnamefont {T.}~\bibnamefont
  {Sato}}, \bibinfo {author} {\bibfnamefont {Y.}~\bibnamefont {Tanaka}},
  \bibinfo {author} {\bibfnamefont {K.}~\bibnamefont {Nakayama}}, \bibinfo
  {author} {\bibfnamefont {S.}~\bibnamefont {Souma}}, \bibinfo {author}
  {\bibfnamefont {T.}~\bibnamefont {Takahashi}}, \bibinfo {author}
  {\bibfnamefont {S.}~\bibnamefont {Sasaki}}, \bibinfo {author} {\bibfnamefont
  {Z.}~\bibnamefont {Ren}}, \bibinfo {author} {\bibfnamefont {A.~A.}\
  \bibnamefont {Taskin}}, \bibinfo {author} {\bibfnamefont {K.}~\bibnamefont
  {Segawa}},\ and\ \bibinfo {author} {\bibfnamefont {Y.}~\bibnamefont {Ando}},\
  }\href {https://doi.org/10.1103/PhysRevLett.110.206804} {\bibfield  {journal}
  {\bibinfo  {journal} {Phys. Rev. Lett.}\ }\textbf {\bibinfo {volume} {110}},\
  \bibinfo {pages} {206804} (\bibinfo {year} {2013})}\BibitemShut {NoStop}%
\bibitem [{\citenamefont {Wray}\ \emph {et~al.}(2010)\citenamefont {Wray},
  \citenamefont {Xu}, \citenamefont {Xia}, \citenamefont {Hor}, \citenamefont
  {Qian}, \citenamefont {Fedorov}, \citenamefont {Lin}, \citenamefont {Bansil},
  \citenamefont {Cava},\ and\ \citenamefont {Hasan}}]{Wray-CBS-ARPES}%
  \BibitemOpen
  \bibfield  {author} {\bibinfo {author} {\bibfnamefont {L.}~\bibnamefont
  {Wray}}, \bibinfo {author} {\bibfnamefont {S.}~\bibnamefont {Xu}}, \bibinfo
  {author} {\bibfnamefont {Y.}~\bibnamefont {Xia}}, \bibinfo {author}
  {\bibfnamefont {Y.}~\bibnamefont {Hor}}, \bibinfo {author} {\bibfnamefont
  {D.}~\bibnamefont {Qian}}, \bibinfo {author} {\bibfnamefont {A.}~\bibnamefont
  {Fedorov}}, \bibinfo {author} {\bibfnamefont {H.}~\bibnamefont {Lin}},
  \bibinfo {author} {\bibfnamefont {A.}~\bibnamefont {Bansil}}, \bibinfo
  {author} {\bibfnamefont {R.}~\bibnamefont {Cava}},\ and\ \bibinfo {author}
  {\bibfnamefont {M.}~\bibnamefont {Hasan}},\ }\href
  {https://doi.org/10.1038/nphys1762} {\bibfield  {journal} {\bibinfo
  {journal} {Nature Physics}\ }\textbf {\bibinfo {volume} {6}},\ \bibinfo
  {pages} {855} (\bibinfo {year} {2010})}\BibitemShut {NoStop}%
\bibitem [{\citenamefont {Neupane}\ \emph {et~al.}(2016)\citenamefont
  {Neupane}, \citenamefont {Ishida}, \citenamefont {Sankar}, \citenamefont
  {Zhu}, \citenamefont {Sanchez}, \citenamefont {Belopolski}, \citenamefont
  {Xu}, \citenamefont {Alidoust}, \citenamefont {Hosen}, \citenamefont {Shin},
  \citenamefont {Chou}, \citenamefont {Hasan},\ and\ \citenamefont
  {Durakiewicz}}]{Neupane-SBS-ARPES}%
  \BibitemOpen
  \bibfield  {author} {\bibinfo {author} {\bibfnamefont {M.}~\bibnamefont
  {Neupane}}, \bibinfo {author} {\bibfnamefont {Y.}~\bibnamefont {Ishida}},
  \bibinfo {author} {\bibfnamefont {R.}~\bibnamefont {Sankar}}, \bibinfo
  {author} {\bibfnamefont {J.-X.}\ \bibnamefont {Zhu}}, \bibinfo {author}
  {\bibfnamefont {D.~S.}\ \bibnamefont {Sanchez}}, \bibinfo {author}
  {\bibfnamefont {I.}~\bibnamefont {Belopolski}}, \bibinfo {author}
  {\bibfnamefont {S.-Y.}\ \bibnamefont {Xu}}, \bibinfo {author} {\bibfnamefont
  {N.}~\bibnamefont {Alidoust}}, \bibinfo {author} {\bibfnamefont {M.~M.}\
  \bibnamefont {Hosen}}, \bibinfo {author} {\bibfnamefont {S.}~\bibnamefont
  {Shin}}, \bibinfo {author} {\bibfnamefont {F.}~\bibnamefont {Chou}}, \bibinfo
  {author} {\bibfnamefont {M.~Z.}\ \bibnamefont {Hasan}},\ and\ \bibinfo
  {author} {\bibfnamefont {T.}~\bibnamefont {Durakiewicz}},\ }\href
  {https://doi.org/10.1038/srep22557} {\bibfield  {journal} {\bibinfo
  {journal} {Scientific Reports}\ }\textbf {\bibinfo {volume} {6}},\ \bibinfo
  {pages} {22557} (\bibinfo {year} {2016})}\BibitemShut {NoStop}%
\bibitem [{\citenamefont {Yonezawa}(2018)}]{Yonezawa-Bi2Se3-review}%
  \BibitemOpen
  \bibfield  {author} {\bibinfo {author} {\bibfnamefont {S.}~\bibnamefont
  {Yonezawa}},\ }\href {https://doi.org/10.3390/condmat4010002} {\bibfield
  {journal} {\bibinfo  {journal} {Condensed Matter}\ }\textbf {\bibinfo
  {volume} {4}},\ \bibinfo {pages} {2} (\bibinfo {year} {2018})}\BibitemShut
  {NoStop}%
\bibitem [{\citenamefont {Fu}\ and\ \citenamefont {Berg}(2010)}]{FuBerg-CBS}%
  \BibitemOpen
  \bibfield  {author} {\bibinfo {author} {\bibfnamefont {L.}~\bibnamefont
  {Fu}}\ and\ \bibinfo {author} {\bibfnamefont {E.}~\bibnamefont {Berg}},\
  }\href {https://doi.org/10.1103/PhysRevLett.105.097001} {\bibfield  {journal}
  {\bibinfo  {journal} {Phys. Rev. Lett.}\ }\textbf {\bibinfo {volume} {105}},\
  \bibinfo {pages} {097001} (\bibinfo {year} {2010})}\BibitemShut {NoStop}%
\bibitem [{\citenamefont {Fu}(2014)}]{FuCBS}%
  \BibitemOpen
  \bibfield  {author} {\bibinfo {author} {\bibfnamefont {L.}~\bibnamefont
  {Fu}},\ }\href {https://doi.org/10.1103/PhysRevB.90.100509} {\bibfield
  {journal} {\bibinfo  {journal} {Phys. Rev. B}\ }\textbf {\bibinfo {volume}
  {90}},\ \bibinfo {pages} {100509(R)} (\bibinfo {year} {2014})}\BibitemShut
  {NoStop}%
\bibitem [{\citenamefont {Sasaki}\ \emph {et~al.}(2012)\citenamefont {Sasaki},
  \citenamefont {Ren}, \citenamefont {Taskin}, \citenamefont {Segawa},
  \citenamefont {Fu},\ and\ \citenamefont {Ando}}]{Sasaki-SnInTe-TSC-data}%
  \BibitemOpen
  \bibfield  {author} {\bibinfo {author} {\bibfnamefont {S.}~\bibnamefont
  {Sasaki}}, \bibinfo {author} {\bibfnamefont {Z.}~\bibnamefont {Ren}},
  \bibinfo {author} {\bibfnamefont {A.~A.}\ \bibnamefont {Taskin}}, \bibinfo
  {author} {\bibfnamefont {K.}~\bibnamefont {Segawa}}, \bibinfo {author}
  {\bibfnamefont {L.}~\bibnamefont {Fu}},\ and\ \bibinfo {author}
  {\bibfnamefont {Y.}~\bibnamefont {Ando}},\ }\href
  {https://doi.org/10.1103/PhysRevLett.109.217004} {\bibfield  {journal}
  {\bibinfo  {journal} {Phys. Rev. Lett.}\ }\textbf {\bibinfo {volume} {109}},\
  \bibinfo {pages} {217004} (\bibinfo {year} {2012})}\BibitemShut {NoStop}%
\bibitem [{\citenamefont {Schmidt}\ and\ \citenamefont
  {Srivastava}(2019)}]{Schmidt-SnInTe-topological}%
  \BibitemOpen
  \bibfield  {author} {\bibinfo {author} {\bibfnamefont {T.~M.}\ \bibnamefont
  {Schmidt}}\ and\ \bibinfo {author} {\bibfnamefont {G.~P.}\ \bibnamefont
  {Srivastava}},\ }\Eprint {https://arxiv.org/abs/1908.05967}
  {arXiv:1908.05967}  (\bibinfo {year} {2019})\BibitemShut {NoStop}%
\bibitem [{\citenamefont {Beenakker}(2013)}]{ProximityBeenakker}%
  \BibitemOpen
  \bibfield  {author} {\bibinfo {author} {\bibfnamefont {C.}~\bibnamefont
  {Beenakker}},\ }\href
  {https://doi.org/10.1146/annurev-conmatphys-030212-184337} {\bibfield
  {journal} {\bibinfo  {journal} {Annual Review of Condensed Matter Physics}\
  }\textbf {\bibinfo {volume} {4}},\ \bibinfo {pages} {113} (\bibinfo {year}
  {2013})}\BibitemShut {NoStop}%
\bibitem [{\citenamefont {Aggarwal}\ \emph {et~al.}(2016)\citenamefont
  {Aggarwal}, \citenamefont {Banik}, \citenamefont {Anand}, \citenamefont
  {Waghmare}, \citenamefont {Biswas},\ and\ \citenamefont
  {Sheet}}]{Aggarwal-SnTe-R3m}%
  \BibitemOpen
  \bibfield  {author} {\bibinfo {author} {\bibfnamefont {L.}~\bibnamefont
  {Aggarwal}}, \bibinfo {author} {\bibfnamefont {A.}~\bibnamefont {Banik}},
  \bibinfo {author} {\bibfnamefont {S.}~\bibnamefont {Anand}}, \bibinfo
  {author} {\bibfnamefont {U.~V.}\ \bibnamefont {Waghmare}}, \bibinfo {author}
  {\bibfnamefont {K.}~\bibnamefont {Biswas}},\ and\ \bibinfo {author}
  {\bibfnamefont {G.}~\bibnamefont {Sheet}},\ }\href
  {https://doi.org/https://doi.org/10.1016/j.jmat.2016.04.001} {\bibfield
  {journal} {\bibinfo  {journal} {Journal of Materiomics}\ }\textbf {\bibinfo
  {volume} {2}},\ \bibinfo {pages} {196} (\bibinfo {year} {2016})},\ \bibinfo
  {note} {special Issue on Advances in Thermoelectric Research}\BibitemShut
  {NoStop}%
\bibitem [{\citenamefont {Hein}(1966)}]{Hein-SnTe-SC}%
  \BibitemOpen
  \bibfield  {author} {\bibinfo {author} {\bibfnamefont {R.~A.}\ \bibnamefont
  {Hein}},\ }\href {https://doi.org/10.1016/0031-9163(66)91080-8} {\bibfield
  {journal} {\bibinfo  {journal} {Phys. Lett.}\ }\textbf {\bibinfo {volume}
  {23}},\ \bibinfo {pages} {435} (\bibinfo {year} {1966})}\BibitemShut
  {NoStop}%
\bibitem [{\citenamefont {Dughaish}(2002)}]{Dughaish-PbTe}%
  \BibitemOpen
  \bibfield  {author} {\bibinfo {author} {\bibfnamefont {Z.~H.}\ \bibnamefont
  {Dughaish}},\ }\href {https://doi.org/10.1016/S0921-4526(02)01187-0}
  {\bibfield  {journal} {\bibinfo  {journal} {Physica B: Condensed Matter}\
  }\textbf {\bibinfo {volume} {322}},\ \bibinfo {pages} {205} (\bibinfo {year}
  {2002})}\BibitemShut {NoStop}%
\bibitem [{\citenamefont {LaLonde}\ \emph {et~al.}(2011)\citenamefont
  {LaLonde}, \citenamefont {Pei}, \citenamefont {Wang},\ and\ \citenamefont
  {{Jeffrey Snyder}}}]{Lalonde-PbTe}%
  \BibitemOpen
  \bibfield  {author} {\bibinfo {author} {\bibfnamefont {A.~D.}\ \bibnamefont
  {LaLonde}}, \bibinfo {author} {\bibfnamefont {Y.}~\bibnamefont {Pei}},
  \bibinfo {author} {\bibfnamefont {H.}~\bibnamefont {Wang}},\ and\ \bibinfo
  {author} {\bibfnamefont {G.}~\bibnamefont {{Jeffrey Snyder}}},\ }\href
  {https://doi.org/https://doi.org/10.1016/S1369-7021(11)70278-4} {\bibfield
  {journal} {\bibinfo  {journal} {Materials Today}\ }\textbf {\bibinfo {volume}
  {14}},\ \bibinfo {pages} {526} (\bibinfo {year} {2011})}\BibitemShut
  {NoStop}%
\bibitem [{\citenamefont {Kang}\ \emph {et~al.}(1987)\citenamefont {Kang},
  \citenamefont {Maps}, \citenamefont {Berkley}, \citenamefont {Jaeger},
  \citenamefont {Goldman},\ and\ \citenamefont {Partin}}]{Kang-PbTeTl}%
  \BibitemOpen
  \bibfield  {author} {\bibinfo {author} {\bibfnamefont {J.~H.}\ \bibnamefont
  {Kang}}, \bibinfo {author} {\bibfnamefont {J.}~\bibnamefont {Maps}}, \bibinfo
  {author} {\bibfnamefont {D.~D.}\ \bibnamefont {Berkley}}, \bibinfo {author}
  {\bibfnamefont {H.~M.}\ \bibnamefont {Jaeger}}, \bibinfo {author}
  {\bibfnamefont {A.~M.}\ \bibnamefont {Goldman}},\ and\ \bibinfo {author}
  {\bibfnamefont {D.~L.}\ \bibnamefont {Partin}},\ }\href
  {https://doi.org/10.1103/PhysRevB.36.2280} {\bibfield  {journal} {\bibinfo
  {journal} {Phys. Rev. B}\ }\textbf {\bibinfo {volume} {36}},\ \bibinfo
  {pages} {2280} (\bibinfo {year} {1987})}\BibitemShut {NoStop}%
\bibitem [{\citenamefont {Hogg}\ and\ \citenamefont
  {Sutherland}(1976)}]{Hogg-InTe-structure}%
  \BibitemOpen
  \bibfield  {author} {\bibinfo {author} {\bibfnamefont {J.~H.~C.}\
  \bibnamefont {Hogg}}\ and\ \bibinfo {author} {\bibfnamefont {H.~H.}\
  \bibnamefont {Sutherland}},\ }\href
  {https://doi.org/10.1107/S056774087600856X} {\bibfield  {journal} {\bibinfo
  {journal} {Acta Crystallographica Section B}\ }\textbf {\bibinfo {volume}
  {32}},\ \bibinfo {pages} {2689} (\bibinfo {year} {1976})}\BibitemShut
  {NoStop}%
\bibitem [{\citenamefont {Tittmann}\ \emph {et~al.}(1964)\citenamefont
  {Tittmann}, \citenamefont {Darnell}, \citenamefont {B\"ommel},\ and\
  \citenamefont {Libby}}]{Tittmann-InTe}%
  \BibitemOpen
  \bibfield  {author} {\bibinfo {author} {\bibfnamefont {B.~R.}\ \bibnamefont
  {Tittmann}}, \bibinfo {author} {\bibfnamefont {A.~J.}\ \bibnamefont
  {Darnell}}, \bibinfo {author} {\bibfnamefont {H.~E.}\ \bibnamefont
  {B\"ommel}},\ and\ \bibinfo {author} {\bibfnamefont {h.~F.}\ \bibnamefont
  {Libby}},\ }\href {https://doi.org/10.1103/PhysRev.135.A1460} {\bibfield
  {journal} {\bibinfo  {journal} {Phys. Rev.}\ }\textbf {\bibinfo {volume}
  {135}},\ \bibinfo {pages} {A1460} (\bibinfo {year} {1964})}\BibitemShut
  {NoStop}%
\bibitem [{\citenamefont {Schmidt}\ and\ \citenamefont
  {Srivastava}(2020)}]{Schmidt-CMS-SnInTe}%
  \BibitemOpen
  \bibfield  {author} {\bibinfo {author} {\bibfnamefont {T.~M.}\ \bibnamefont
  {Schmidt}}\ and\ \bibinfo {author} {\bibfnamefont {G.}~\bibnamefont
  {Srivastava}},\ }\href
  {https://doi.org/https://doi.org/10.1016/j.commatsci.2020.109777} {\bibfield
  {journal} {\bibinfo  {journal} {Computational Materials Science}\ }\textbf
  {\bibinfo {volume} {182}},\ \bibinfo {pages} {109777} (\bibinfo {year}
  {2020})}\BibitemShut {NoStop}%
\bibitem [{\citenamefont {Novak}\ \emph {et~al.}(2013)\citenamefont {Novak},
  \citenamefont {Sasaki}, \citenamefont {Kriener}, \citenamefont {Segawa},\
  and\ \citenamefont {Ando}}]{Novak-SnInTe-disorder}%
  \BibitemOpen
  \bibfield  {author} {\bibinfo {author} {\bibfnamefont {M.}~\bibnamefont
  {Novak}}, \bibinfo {author} {\bibfnamefont {S.}~\bibnamefont {Sasaki}},
  \bibinfo {author} {\bibfnamefont {M.}~\bibnamefont {Kriener}}, \bibinfo
  {author} {\bibfnamefont {K.}~\bibnamefont {Segawa}},\ and\ \bibinfo {author}
  {\bibfnamefont {Y.}~\bibnamefont {Ando}},\ }\href
  {https://doi.org/10.1103/PhysRevB.88.140502} {\bibfield  {journal} {\bibinfo
  {journal} {Phys. Rev. B}\ }\textbf {\bibinfo {volume} {88}},\ \bibinfo
  {pages} {140502(R)} (\bibinfo {year} {2013})}\BibitemShut {NoStop}%
\bibitem [{\citenamefont {Zhong}\ \emph {et~al.}(2013)\citenamefont {Zhong},
  \citenamefont {Schneeloch}, \citenamefont {Shi}, \citenamefont {Xu},
  \citenamefont {Zhang}, \citenamefont {Tranquada}, \citenamefont {Li},\ and\
  \citenamefont {Gu}}]{Zhong-SnInTe-growth}%
  \BibitemOpen
  \bibfield  {author} {\bibinfo {author} {\bibfnamefont {R.~D.}\ \bibnamefont
  {Zhong}}, \bibinfo {author} {\bibfnamefont {J.~A.}\ \bibnamefont
  {Schneeloch}}, \bibinfo {author} {\bibfnamefont {X.~Y.}\ \bibnamefont {Shi}},
  \bibinfo {author} {\bibfnamefont {Z.~J.}\ \bibnamefont {Xu}}, \bibinfo
  {author} {\bibfnamefont {C.}~\bibnamefont {Zhang}}, \bibinfo {author}
  {\bibfnamefont {J.~M.}\ \bibnamefont {Tranquada}}, \bibinfo {author}
  {\bibfnamefont {Q.}~\bibnamefont {Li}},\ and\ \bibinfo {author}
  {\bibfnamefont {G.~D.}\ \bibnamefont {Gu}},\ }\href
  {https://doi.org/10.1103/PhysRevB.88.020505} {\bibfield  {journal} {\bibinfo
  {journal} {Phys. Rev. B}\ }\textbf {\bibinfo {volume} {88}},\ \bibinfo
  {pages} {020505(R)} (\bibinfo {year} {2013})}\BibitemShut {NoStop}%
\bibitem [{\citenamefont {Xu}\ \emph {et~al.}(2012)\citenamefont {Xu},
  \citenamefont {Liu}, \citenamefont {Alidoust}, \citenamefont {Neupane},
  \citenamefont {Qian}, \citenamefont {Belopolski}, \citenamefont {Denlinger},
  \citenamefont {Wang}, \citenamefont {Lin}, \citenamefont {Wray},
  \citenamefont {Landolt}, \citenamefont {Slomski}, \citenamefont {Dil},
  \citenamefont {Marcinkova}, \citenamefont {Morosan}, \citenamefont {Gibson},
  \citenamefont {Sankar}, \citenamefont {Chou}, \citenamefont {Cava},
  \citenamefont {Bansil},\ and\ \citenamefont {Hasan}}]{Xu-NatComm-PbSnTe}%
  \BibitemOpen
  \bibfield  {author} {\bibinfo {author} {\bibfnamefont {S.-Y.}\ \bibnamefont
  {Xu}}, \bibinfo {author} {\bibfnamefont {C.}~\bibnamefont {Liu}}, \bibinfo
  {author} {\bibfnamefont {N.}~\bibnamefont {Alidoust}}, \bibinfo {author}
  {\bibfnamefont {M.}~\bibnamefont {Neupane}}, \bibinfo {author} {\bibfnamefont
  {D.}~\bibnamefont {Qian}}, \bibinfo {author} {\bibfnamefont {I.}~\bibnamefont
  {Belopolski}}, \bibinfo {author} {\bibfnamefont {J.~D.}\ \bibnamefont
  {Denlinger}}, \bibinfo {author} {\bibfnamefont {Y.~J.}\ \bibnamefont {Wang}},
  \bibinfo {author} {\bibfnamefont {H.}~\bibnamefont {Lin}}, \bibinfo {author}
  {\bibfnamefont {L.~A.}\ \bibnamefont {Wray}}, \bibinfo {author}
  {\bibfnamefont {G.}~\bibnamefont {Landolt}}, \bibinfo {author} {\bibfnamefont
  {B.}~\bibnamefont {Slomski}}, \bibinfo {author} {\bibfnamefont {J.~H.}\
  \bibnamefont {Dil}}, \bibinfo {author} {\bibfnamefont {A.}~\bibnamefont
  {Marcinkova}}, \bibinfo {author} {\bibfnamefont {E.}~\bibnamefont {Morosan}},
  \bibinfo {author} {\bibfnamefont {Q.}~\bibnamefont {Gibson}}, \bibinfo
  {author} {\bibfnamefont {R.}~\bibnamefont {Sankar}}, \bibinfo {author}
  {\bibfnamefont {F.~C.}\ \bibnamefont {Chou}}, \bibinfo {author}
  {\bibfnamefont {R.~J.}\ \bibnamefont {Cava}}, \bibinfo {author}
  {\bibfnamefont {A.}~\bibnamefont {Bansil}},\ and\ \bibinfo {author}
  {\bibfnamefont {M.~Z.}\ \bibnamefont {Hasan}},\ }\href
  {https://doi.org/10.1038/ncomms2191} {\bibfield  {journal} {\bibinfo
  {journal} {Nature Communications}\ }\textbf {\bibinfo {volume} {3}},\
  \bibinfo {pages} {1192} (\bibinfo {year} {2012})}\BibitemShut {NoStop}%
\bibitem [{\citenamefont {Tanaka}\ \emph {et~al.}(2013)\citenamefont {Tanaka},
  \citenamefont {Sato}, \citenamefont {Nakayama}, \citenamefont {Souma},
  \citenamefont {Takahashi}, \citenamefont {Ren}, \citenamefont {Novak},
  \citenamefont {Segawa},\ and\ \citenamefont {Ando}}]{Tanaka-PRB-PbSnTe}%
  \BibitemOpen
  \bibfield  {author} {\bibinfo {author} {\bibfnamefont {Y.}~\bibnamefont
  {Tanaka}}, \bibinfo {author} {\bibfnamefont {T.}~\bibnamefont {Sato}},
  \bibinfo {author} {\bibfnamefont {K.}~\bibnamefont {Nakayama}}, \bibinfo
  {author} {\bibfnamefont {S.}~\bibnamefont {Souma}}, \bibinfo {author}
  {\bibfnamefont {T.}~\bibnamefont {Takahashi}}, \bibinfo {author}
  {\bibfnamefont {Z.}~\bibnamefont {Ren}}, \bibinfo {author} {\bibfnamefont
  {M.}~\bibnamefont {Novak}}, \bibinfo {author} {\bibfnamefont
  {K.}~\bibnamefont {Segawa}},\ and\ \bibinfo {author} {\bibfnamefont
  {Y.}~\bibnamefont {Ando}},\ }\href
  {https://doi.org/10.1103/PhysRevB.87.155105} {\bibfield  {journal} {\bibinfo
  {journal} {Phys. Rev. B}\ }\textbf {\bibinfo {volume} {87}},\ \bibinfo
  {pages} {155105} (\bibinfo {year} {2013})}\BibitemShut {NoStop}%
\bibitem [{\citenamefont {Ravich}\ and\ \citenamefont
  {Nemov}(2002)}]{Ravich-PbInTe}%
  \BibitemOpen
  \bibfield  {author} {\bibinfo {author} {\bibfnamefont {Y.~I.}\ \bibnamefont
  {Ravich}}\ and\ \bibinfo {author} {\bibfnamefont {S.~A.}\ \bibnamefont
  {Nemov}},\ }\href {https://doi.org/10.1134/1.1434506} {\bibfield  {journal}
  {\bibinfo  {journal} {Semiconductors}\ }\textbf {\bibinfo {volume} {36}},\
  \bibinfo {pages} {1} (\bibinfo {year} {2002})}\BibitemShut {NoStop}%
\bibitem [{\citenamefont {Rosenberg}\ \emph {et~al.}(1964)\citenamefont
  {Rosenberg}, \citenamefont {Wooley}, \citenamefont {Nikolic},\ and\
  \citenamefont {Grierson}}]{Rosenberg-PbInTe}%
  \BibitemOpen
  \bibfield  {author} {\bibinfo {author} {\bibfnamefont {A.~I.}\ \bibnamefont
  {Rosenberg}}, \bibinfo {author} {\bibfnamefont {J.~C.}\ \bibnamefont
  {Wooley}}, \bibinfo {author} {\bibfnamefont {P.}~\bibnamefont {Nikolic}},\
  and\ \bibinfo {author} {\bibfnamefont {R.}~\bibnamefont {Grierson}},\
  }\href@noop {} {\bibfield  {journal} {\bibinfo  {journal} {Transactions of
  the Metallurgical Society of AIME}\ }\textbf {\bibinfo {volume} {230}},\
  \bibinfo {pages} {342} (\bibinfo {year} {1964})}\BibitemShut {NoStop}%
\bibitem [{\citenamefont {Zhong}\ \emph {et~al.}(2017)\citenamefont {Zhong},
  \citenamefont {Schneeloch}, \citenamefont {Li}, \citenamefont {Ku},
  \citenamefont {Tranquada},\ and\ \citenamefont
  {Gu}}]{Zhong-PbSnInTe-phase-diagram}%
  \BibitemOpen
  \bibfield  {author} {\bibinfo {author} {\bibfnamefont {R.}~\bibnamefont
  {Zhong}}, \bibinfo {author} {\bibfnamefont {J.}~\bibnamefont {Schneeloch}},
  \bibinfo {author} {\bibfnamefont {Q.}~\bibnamefont {Li}}, \bibinfo {author}
  {\bibfnamefont {W.}~\bibnamefont {Ku}}, \bibinfo {author} {\bibfnamefont
  {J.}~\bibnamefont {Tranquada}},\ and\ \bibinfo {author} {\bibfnamefont
  {G.}~\bibnamefont {Gu}},\ }\href {https://doi.org/10.3390/cryst7020055}
  {\bibfield  {journal} {\bibinfo  {journal} {Crystals}\ }\textbf {\bibinfo
  {volume} {7}},\ \bibinfo {pages} {55} (\bibinfo {year} {2017})}\BibitemShut
  {NoStop}%
\bibitem [{\citenamefont {{Mikhailin}}\ \emph {et~al.}(2019)\citenamefont
  {{Mikhailin}}, \citenamefont {{Shamshur}}, \citenamefont {{Volkov}},
  \citenamefont {{Chernyaev}},\ and\ \citenamefont
  {{Parfenev}}}]{Mikhailin-PbSnInTe}%
  \BibitemOpen
  \bibfield  {author} {\bibinfo {author} {\bibfnamefont {N.~Y.}\ \bibnamefont
  {{Mikhailin}}}, \bibinfo {author} {\bibfnamefont {D.~V.}\ \bibnamefont
  {{Shamshur}}}, \bibinfo {author} {\bibfnamefont {M.~P.}\ \bibnamefont
  {{Volkov}}}, \bibinfo {author} {\bibfnamefont {A.~V.}\ \bibnamefont
  {{Chernyaev}}},\ and\ \bibinfo {author} {\bibfnamefont {R.~V.}\ \bibnamefont
  {{Parfenev}}},\ }\href {https://doi.org/10.1063/1.5086409} {\bibfield
  {journal} {\bibinfo  {journal} {Low Temperature Physics}\ }\textbf {\bibinfo
  {volume} {45}},\ \bibinfo {pages} {189} (\bibinfo {year} {2019})}\BibitemShut
  {NoStop}%
\bibitem [{\citenamefont {Denisov}\ \emph {et~al.}(2020)\citenamefont
  {Denisov}, \citenamefont {Mikhailin}, \citenamefont {Shamshur},\ and\
  \citenamefont {Parfeniev}}]{Denisov-PbSnInTe}%
  \BibitemOpen
  \bibfield  {author} {\bibinfo {author} {\bibfnamefont {D.}~\bibnamefont
  {Denisov}}, \bibinfo {author} {\bibfnamefont {N.}~\bibnamefont {Mikhailin}},
  \bibinfo {author} {\bibfnamefont {D.}~\bibnamefont {Shamshur}},\ and\
  \bibinfo {author} {\bibfnamefont {R.}~\bibnamefont {Parfeniev}},\ }\href
  {https://doi.org/10.1016/j.physc.2020.1353755} {\bibfield  {journal}
  {\bibinfo  {journal} {Physica C: Superconductivity and its Applications}\
  }\textbf {\bibinfo {volume} {579}},\ \bibinfo {pages} {1353755} (\bibinfo
  {year} {2020})}\BibitemShut {NoStop}%
\bibitem [{\citenamefont {Kobayashi}\ \emph {et~al.}(2018)\citenamefont
  {Kobayashi}, \citenamefont {Ai}, \citenamefont {Jeschke},\ and\ \citenamefont
  {Akimitsu}}]{Kobayashi-HPHT-SnInTe}%
  \BibitemOpen
  \bibfield  {author} {\bibinfo {author} {\bibfnamefont {K.}~\bibnamefont
  {Kobayashi}}, \bibinfo {author} {\bibfnamefont {Y.}~\bibnamefont {Ai}},
  \bibinfo {author} {\bibfnamefont {H.~O.}\ \bibnamefont {Jeschke}},\ and\
  \bibinfo {author} {\bibfnamefont {J.}~\bibnamefont {Akimitsu}},\ }\href
  {https://doi.org/10.1103/PhysRevB.97.104511} {\bibfield  {journal} {\bibinfo
  {journal} {Phys. Rev. B}\ }\textbf {\bibinfo {volume} {97}},\ \bibinfo
  {pages} {104511} (\bibinfo {year} {2018})}\BibitemShut {NoStop}%
\bibitem [{\citenamefont {Katsuno}\ \emph {et~al.}(2020)\citenamefont
  {Katsuno}, \citenamefont {Jha}, \citenamefont {Hoshi}, \citenamefont
  {Sogabe}, \citenamefont {Goto},\ and\ \citenamefont
  {Mizuguchi}}]{Katsuno-HPHT-InPbTe}%
  \BibitemOpen
  \bibfield  {author} {\bibinfo {author} {\bibfnamefont {M.}~\bibnamefont
  {Katsuno}}, \bibinfo {author} {\bibfnamefont {R.}~\bibnamefont {Jha}},
  \bibinfo {author} {\bibfnamefont {K.}~\bibnamefont {Hoshi}}, \bibinfo
  {author} {\bibfnamefont {R.}~\bibnamefont {Sogabe}}, \bibinfo {author}
  {\bibfnamefont {Y.}~\bibnamefont {Goto}},\ and\ \bibinfo {author}
  {\bibfnamefont {Y.}~\bibnamefont {Mizuguchi}},\ }\href
  {https://doi.org/10.3390/condmat5010014} {\bibfield  {journal} {\bibinfo
  {journal} {Condensed Matter}\ }\textbf {\bibinfo {volume} {5}},\ \bibinfo
  {pages} {14} (\bibinfo {year} {2020})}\BibitemShut {NoStop}%
\bibitem [{\citenamefont {Rajaji}\ \emph {et~al.}(2018)\citenamefont {Rajaji},
  \citenamefont {Pal}, \citenamefont {Sarma}, \citenamefont {Joseph},
  \citenamefont {Peter}, \citenamefont {Waghmare},\ and\ \citenamefont
  {Narayana}}]{Rajaji-InTe-TCI}%
  \BibitemOpen
  \bibfield  {author} {\bibinfo {author} {\bibfnamefont {V.}~\bibnamefont
  {Rajaji}}, \bibinfo {author} {\bibfnamefont {K.}~\bibnamefont {Pal}},
  \bibinfo {author} {\bibfnamefont {S.~C.}\ \bibnamefont {Sarma}}, \bibinfo
  {author} {\bibfnamefont {B.}~\bibnamefont {Joseph}}, \bibinfo {author}
  {\bibfnamefont {S.~C.}\ \bibnamefont {Peter}}, \bibinfo {author}
  {\bibfnamefont {U.~V.}\ \bibnamefont {Waghmare}},\ and\ \bibinfo {author}
  {\bibfnamefont {C.}~\bibnamefont {Narayana}},\ }\href
  {https://doi.org/10.1103/PhysRevB.97.155158} {\bibfield  {journal} {\bibinfo
  {journal} {Phys. Rev. B}\ }\textbf {\bibinfo {volume} {97}},\ \bibinfo
  {pages} {155158} (\bibinfo {year} {2018})}\BibitemShut {NoStop}%
\bibitem [{\citenamefont {Du}\ \emph {et~al.}(2015)\citenamefont {Du},
  \citenamefont {Du}, \citenamefont {Fang}, \citenamefont {Yang}, \citenamefont
  {Zhong}, \citenamefont {Schneeloch}, \citenamefont {Gu},\ and\ \citenamefont
  {Wen}}]{Du-PbSnTe-full-gap-STS}%
  \BibitemOpen
  \bibfield  {author} {\bibinfo {author} {\bibfnamefont {G.}~\bibnamefont
  {Du}}, \bibinfo {author} {\bibfnamefont {Z.}~\bibnamefont {Du}}, \bibinfo
  {author} {\bibfnamefont {D.}~\bibnamefont {Fang}}, \bibinfo {author}
  {\bibfnamefont {H.}~\bibnamefont {Yang}}, \bibinfo {author} {\bibfnamefont
  {R.~D.}\ \bibnamefont {Zhong}}, \bibinfo {author} {\bibfnamefont
  {J.}~\bibnamefont {Schneeloch}}, \bibinfo {author} {\bibfnamefont {G.~D.}\
  \bibnamefont {Gu}},\ and\ \bibinfo {author} {\bibfnamefont {H.-H.}\
  \bibnamefont {Wen}},\ }\href {https://doi.org/10.1103/PhysRevB.92.020512}
  {\bibfield  {journal} {\bibinfo  {journal} {Phys. Rev. B}\ }\textbf {\bibinfo
  {volume} {92}},\ \bibinfo {pages} {020512(R)} (\bibinfo {year}
  {2015})}\BibitemShut {NoStop}%
\bibitem [{\citenamefont {Smylie}\ \emph {et~al.}(2018)\citenamefont {Smylie},
  \citenamefont {Claus}, \citenamefont {Kwok}, \citenamefont {Louden},
  \citenamefont {Eskildsen}, \citenamefont {Sefat}, \citenamefont {Zhong},
  \citenamefont {Schneeloch}, \citenamefont {Gu}, \citenamefont {Bokari},
  \citenamefont {Niraula}, \citenamefont {Kayani}, \citenamefont {Dewhurst},
  \citenamefont {Snezhko},\ and\ \citenamefont {Welp}}]{Smylie-SnInTe}%
  \BibitemOpen
  \bibfield  {author} {\bibinfo {author} {\bibfnamefont {M.~P.}\ \bibnamefont
  {Smylie}}, \bibinfo {author} {\bibfnamefont {H.}~\bibnamefont {Claus}},
  \bibinfo {author} {\bibfnamefont {W.-K.}\ \bibnamefont {Kwok}}, \bibinfo
  {author} {\bibfnamefont {E.~R.}\ \bibnamefont {Louden}}, \bibinfo {author}
  {\bibfnamefont {M.~R.}\ \bibnamefont {Eskildsen}}, \bibinfo {author}
  {\bibfnamefont {A.~S.}\ \bibnamefont {Sefat}}, \bibinfo {author}
  {\bibfnamefont {R.~D.}\ \bibnamefont {Zhong}}, \bibinfo {author}
  {\bibfnamefont {J.}~\bibnamefont {Schneeloch}}, \bibinfo {author}
  {\bibfnamefont {G.~D.}\ \bibnamefont {Gu}}, \bibinfo {author} {\bibfnamefont
  {E.}~\bibnamefont {Bokari}}, \bibinfo {author} {\bibfnamefont {P.~M.}\
  \bibnamefont {Niraula}}, \bibinfo {author} {\bibfnamefont {A.}~\bibnamefont
  {Kayani}}, \bibinfo {author} {\bibfnamefont {C.~D.}\ \bibnamefont
  {Dewhurst}}, \bibinfo {author} {\bibfnamefont {A.}~\bibnamefont {Snezhko}},\
  and\ \bibinfo {author} {\bibfnamefont {U.}~\bibnamefont {Welp}},\ }\href
  {https://doi.org/10.1103/PhysRevB.97.024511} {\bibfield  {journal} {\bibinfo
  {journal} {Phys. Rev. B}\ }\textbf {\bibinfo {volume} {97}},\ \bibinfo
  {pages} {024511} (\bibinfo {year} {2018})}\BibitemShut {NoStop}%
\bibitem [{\citenamefont {Smylie}\ \emph {et~al.}(2020)\citenamefont {Smylie},
  \citenamefont {Kobayashi}, \citenamefont {Takahashi}, \citenamefont
  {Chaparro}, \citenamefont {Snezhko}, \citenamefont {Kwok},\ and\
  \citenamefont {Welp}}]{Smylie-SnInTe-HPHT}%
  \BibitemOpen
  \bibfield  {author} {\bibinfo {author} {\bibfnamefont {M.~P.}\ \bibnamefont
  {Smylie}}, \bibinfo {author} {\bibfnamefont {K.}~\bibnamefont {Kobayashi}},
  \bibinfo {author} {\bibfnamefont {T.}~\bibnamefont {Takahashi}}, \bibinfo
  {author} {\bibfnamefont {C.}~\bibnamefont {Chaparro}}, \bibinfo {author}
  {\bibfnamefont {A.}~\bibnamefont {Snezhko}}, \bibinfo {author} {\bibfnamefont
  {W.-K.}\ \bibnamefont {Kwok}},\ and\ \bibinfo {author} {\bibfnamefont
  {U.}~\bibnamefont {Welp}},\ }\href
  {https://doi.org/10.1103/PhysRevB.101.094513} {\bibfield  {journal} {\bibinfo
   {journal} {Phys. Rev. B}\ }\textbf {\bibinfo {volume} {101}},\ \bibinfo
  {pages} {094513} (\bibinfo {year} {2020})}\BibitemShut {NoStop}%
\bibitem [{\citenamefont {Prozorov}\ and\ \citenamefont
  {Giannetta}(2006)}]{ProzorovTDO}%
  \BibitemOpen
  \bibfield  {author} {\bibinfo {author} {\bibfnamefont {R.}~\bibnamefont
  {Prozorov}}\ and\ \bibinfo {author} {\bibfnamefont {R.~W.}\ \bibnamefont
  {Giannetta}},\ }\href {https://doi.org/10.1088/0953-2048/19/8/r01} {\bibfield
   {journal} {\bibinfo  {journal} {Superconductor Science and Technology}\
  }\textbf {\bibinfo {volume} {19}},\ \bibinfo {pages} {R41} (\bibinfo {year}
  {2006})}\BibitemShut {NoStop}%
\bibitem [{\citenamefont {Smylie}\ \emph {et~al.}(2016)\citenamefont {Smylie},
  \citenamefont {Claus}, \citenamefont {Welp}, \citenamefont {Kwok},
  \citenamefont {Qiu}, \citenamefont {Hor},\ and\ \citenamefont
  {Snezhko}}]{SmylieTDO1}%
  \BibitemOpen
  \bibfield  {author} {\bibinfo {author} {\bibfnamefont {M.~P.}\ \bibnamefont
  {Smylie}}, \bibinfo {author} {\bibfnamefont {H.}~\bibnamefont {Claus}},
  \bibinfo {author} {\bibfnamefont {U.}~\bibnamefont {Welp}}, \bibinfo {author}
  {\bibfnamefont {W.-K.}\ \bibnamefont {Kwok}}, \bibinfo {author}
  {\bibfnamefont {Y.}~\bibnamefont {Qiu}}, \bibinfo {author} {\bibfnamefont
  {Y.~S.}\ \bibnamefont {Hor}},\ and\ \bibinfo {author} {\bibfnamefont
  {A.}~\bibnamefont {Snezhko}},\ }\href
  {https://doi.org/10.1103/PhysRevB.94.180510} {\bibfield  {journal} {\bibinfo
  {journal} {Phys. Rev. B}\ }\textbf {\bibinfo {volume} {94}},\ \bibinfo
  {pages} {180510(R)} (\bibinfo {year} {2016})}\BibitemShut {NoStop}%
\bibitem [{\citenamefont {Carrington}\ \emph {et~al.}(1999)\citenamefont
  {Carrington}, \citenamefont {Giannetta}, \citenamefont {Kim},\ and\
  \citenamefont {Giapintzakis}}]{Carrington-TDO-1999}%
  \BibitemOpen
  \bibfield  {author} {\bibinfo {author} {\bibfnamefont {A.}~\bibnamefont
  {Carrington}}, \bibinfo {author} {\bibfnamefont {R.~W.}\ \bibnamefont
  {Giannetta}}, \bibinfo {author} {\bibfnamefont {J.~T.}\ \bibnamefont {Kim}},\
  and\ \bibinfo {author} {\bibfnamefont {J.}~\bibnamefont {Giapintzakis}},\
  }\href {https://doi.org/10.1103/PhysRevB.59.R14173} {\bibfield  {journal}
  {\bibinfo  {journal} {Phys. Rev. B}\ }\textbf {\bibinfo {volume} {59}},\
  \bibinfo {pages} {R14173} (\bibinfo {year} {1999})}\BibitemShut {NoStop}%
\bibitem [{\citenamefont {Prozorov}\ \emph {et~al.}(2000)\citenamefont
  {Prozorov}, \citenamefont {Giannetta}, \citenamefont {Carrington},
  \citenamefont {Fournier}, \citenamefont {Greene}, \citenamefont {Guptasarma},
  \citenamefont {Hinks},\ and\ \citenamefont
  {Banks}}]{Prozorov-Giannetta-APL-2000}%
  \BibitemOpen
  \bibfield  {author} {\bibinfo {author} {\bibfnamefont {R.}~\bibnamefont
  {Prozorov}}, \bibinfo {author} {\bibfnamefont {R.}~\bibnamefont {Giannetta}},
  \bibinfo {author} {\bibfnamefont {A.}~\bibnamefont {Carrington}}, \bibinfo
  {author} {\bibfnamefont {P.}~\bibnamefont {Fournier}}, \bibinfo {author}
  {\bibfnamefont {R.}~\bibnamefont {Greene}}, \bibinfo {author} {\bibfnamefont
  {P.}~\bibnamefont {Guptasarma}}, \bibinfo {author} {\bibfnamefont
  {D.}~\bibnamefont {Hinks}},\ and\ \bibinfo {author} {\bibfnamefont
  {A.}~\bibnamefont {Banks}},\ }\href {https://doi.org/10.1063/1.1328362}
  {\bibfield  {journal} {\bibinfo  {journal} {Applied Physics Letters}\
  }\textbf {\bibinfo {volume} {77}},\ \bibinfo {pages} {4202} (\bibinfo {year}
  {2000})}\BibitemShut {NoStop}%
\bibitem [{\citenamefont {Manzano}\ \emph {et~al.}(2002)\citenamefont
  {Manzano}, \citenamefont {Carrington}, \citenamefont {Hussey}, \citenamefont
  {Lee}, \citenamefont {Yamamoto},\ and\ \citenamefont
  {Tajima}}]{AnisotropicGap1}%
  \BibitemOpen
  \bibfield  {author} {\bibinfo {author} {\bibfnamefont {F.}~\bibnamefont
  {Manzano}}, \bibinfo {author} {\bibfnamefont {A.}~\bibnamefont {Carrington}},
  \bibinfo {author} {\bibfnamefont {N.~E.}\ \bibnamefont {Hussey}}, \bibinfo
  {author} {\bibfnamefont {S.}~\bibnamefont {Lee}}, \bibinfo {author}
  {\bibfnamefont {A.}~\bibnamefont {Yamamoto}},\ and\ \bibinfo {author}
  {\bibfnamefont {S.}~\bibnamefont {Tajima}},\ }\href
  {https://doi.org/10.1103/PhysRevLett.88.047002} {\bibfield  {journal}
  {\bibinfo  {journal} {Phys. Rev. Lett.}\ }\textbf {\bibinfo {volume} {88}},\
  \bibinfo {pages} {047002} (\bibinfo {year} {2002})}\BibitemShut {NoStop}%
\bibitem [{\citenamefont {Fletcher}\ \emph {et~al.}(2007)\citenamefont
  {Fletcher}, \citenamefont {Carrington}, \citenamefont {Diener}, \citenamefont
  {Rodi\`ere}, \citenamefont {Brison}, \citenamefont {Prozorov}, \citenamefont
  {Olheiser},\ and\ \citenamefont {Giannetta}}]{AnisotropicGap2}%
  \BibitemOpen
  \bibfield  {author} {\bibinfo {author} {\bibfnamefont {J.~D.}\ \bibnamefont
  {Fletcher}}, \bibinfo {author} {\bibfnamefont {A.}~\bibnamefont
  {Carrington}}, \bibinfo {author} {\bibfnamefont {P.}~\bibnamefont {Diener}},
  \bibinfo {author} {\bibfnamefont {P.}~\bibnamefont {Rodi\`ere}}, \bibinfo
  {author} {\bibfnamefont {J.~P.}\ \bibnamefont {Brison}}, \bibinfo {author}
  {\bibfnamefont {R.}~\bibnamefont {Prozorov}}, \bibinfo {author}
  {\bibfnamefont {T.}~\bibnamefont {Olheiser}},\ and\ \bibinfo {author}
  {\bibfnamefont {R.~W.}\ \bibnamefont {Giannetta}},\ }\href
  {https://doi.org/10.1103/PhysRevLett.98.057003} {\bibfield  {journal}
  {\bibinfo  {journal} {Phys. Rev. Lett.}\ }\textbf {\bibinfo {volume} {98}},\
  \bibinfo {pages} {057003} (\bibinfo {year} {2007})}\BibitemShut {NoStop}%
\bibitem [{\citenamefont {Marsiglio}\ and\ \citenamefont
  {Carbotte}(2003)}]{AnisotropicGap3}%
  \BibitemOpen
  \bibfield  {author} {\bibinfo {author} {\bibfnamefont {F.}~\bibnamefont
  {Marsiglio}}\ and\ \bibinfo {author} {\bibfnamefont {J.~P.}\ \bibnamefont
  {Carbotte}},\ }\href@noop {} {\emph {\bibinfo {title} {The Physics of
  Superconductivity}}},\ Vol.~\bibinfo {volume} {1}\ (\bibinfo  {publisher}
  {Springer},\ \bibinfo {address} {Berlin},\ \bibinfo {year} {2003})\
  p.~\bibinfo {pages} {73}\BibitemShut {NoStop}%
\bibitem [{\citenamefont {Ran}\ \emph {et~al.}(2018)\citenamefont {Ran},
  \citenamefont {Zhong}, \citenamefont {Chen}, \citenamefont {Gan},
  \citenamefont {Wang}, \citenamefont {Winn}, \citenamefont {Christianson},
  \citenamefont {Li}, \citenamefont {Ma}, \citenamefont {Bao}, \citenamefont
  {Cai}, \citenamefont {Xu}, \citenamefont {Tranquada}, \citenamefont {Gu},
  \citenamefont {Sun},\ and\ \citenamefont {Wen}}]{Ran-PbSnInTe}%
  \BibitemOpen
  \bibfield  {author} {\bibinfo {author} {\bibfnamefont {K.}~\bibnamefont
  {Ran}}, \bibinfo {author} {\bibfnamefont {R.}~\bibnamefont {Zhong}}, \bibinfo
  {author} {\bibfnamefont {T.}~\bibnamefont {Chen}}, \bibinfo {author}
  {\bibfnamefont {Y.}~\bibnamefont {Gan}}, \bibinfo {author} {\bibfnamefont
  {J.}~\bibnamefont {Wang}}, \bibinfo {author} {\bibfnamefont {B.~L.}\
  \bibnamefont {Winn}}, \bibinfo {author} {\bibfnamefont {A.~D.}\ \bibnamefont
  {Christianson}}, \bibinfo {author} {\bibfnamefont {S.}~\bibnamefont {Li}},
  \bibinfo {author} {\bibfnamefont {Z.}~\bibnamefont {Ma}}, \bibinfo {author}
  {\bibfnamefont {S.}~\bibnamefont {Bao}}, \bibinfo {author} {\bibfnamefont
  {Z.}~\bibnamefont {Cai}}, \bibinfo {author} {\bibfnamefont {G.}~\bibnamefont
  {Xu}}, \bibinfo {author} {\bibfnamefont {J.~M.}\ \bibnamefont {Tranquada}},
  \bibinfo {author} {\bibfnamefont {G.}~\bibnamefont {Gu}}, \bibinfo {author}
  {\bibfnamefont {J.}~\bibnamefont {Sun}},\ and\ \bibinfo {author}
  {\bibfnamefont {J.}~\bibnamefont {Wen}},\ }\href
  {https://doi.org/10.1103/PhysRevB.97.220502} {\bibfield  {journal} {\bibinfo
  {journal} {Phys. Rev. B}\ }\textbf {\bibinfo {volume} {97}},\ \bibinfo
  {pages} {220502(R)} (\bibinfo {year} {2018})}\BibitemShut {NoStop}%
\bibitem [{\citenamefont {Sapkota}\ \emph {et~al.}(2020)\citenamefont
  {Sapkota}, \citenamefont {Li}, \citenamefont {Winn}, \citenamefont
  {Podlesnyak}, \citenamefont {Xu}, \citenamefont {Xu}, \citenamefont {Ran},
  \citenamefont {Chen}, \citenamefont {Sun}, \citenamefont {Wen}, \citenamefont
  {Wu}, \citenamefont {Yang}, \citenamefont {Li}, \citenamefont {Gu},\ and\
  \citenamefont {Tranquada}}]{Sapkota-PbSnInTe}%
  \BibitemOpen
  \bibfield  {author} {\bibinfo {author} {\bibfnamefont {A.}~\bibnamefont
  {Sapkota}}, \bibinfo {author} {\bibfnamefont {Y.}~\bibnamefont {Li}},
  \bibinfo {author} {\bibfnamefont {B.~L.}\ \bibnamefont {Winn}}, \bibinfo
  {author} {\bibfnamefont {A.}~\bibnamefont {Podlesnyak}}, \bibinfo {author}
  {\bibfnamefont {G.}~\bibnamefont {Xu}}, \bibinfo {author} {\bibfnamefont
  {Z.}~\bibnamefont {Xu}}, \bibinfo {author} {\bibfnamefont {K.}~\bibnamefont
  {Ran}}, \bibinfo {author} {\bibfnamefont {T.}~\bibnamefont {Chen}}, \bibinfo
  {author} {\bibfnamefont {J.}~\bibnamefont {Sun}}, \bibinfo {author}
  {\bibfnamefont {J.}~\bibnamefont {Wen}}, \bibinfo {author} {\bibfnamefont
  {L.}~\bibnamefont {Wu}}, \bibinfo {author} {\bibfnamefont {J.}~\bibnamefont
  {Yang}}, \bibinfo {author} {\bibfnamefont {Q.}~\bibnamefont {Li}}, \bibinfo
  {author} {\bibfnamefont {G.~D.}\ \bibnamefont {Gu}},\ and\ \bibinfo {author}
  {\bibfnamefont {J.~M.}\ \bibnamefont {Tranquada}},\ }\href
  {https://doi.org/10.1103/PhysRevB.102.104511} {\bibfield  {journal} {\bibinfo
   {journal} {Phys. Rev. B}\ }\textbf {\bibinfo {volume} {102}},\ \bibinfo
  {pages} {104511} (\bibinfo {year} {2020})}\BibitemShut {NoStop}%
\end{thebibliography}%

\end{document}